\documentclass[iop]{emulateapj}

\newcommand\lta{\mathrel{\hbox{\raise 0.6 ex \hbox{$<$}\kern
                   -1.8 ex\lower .5 ex\hbox{$\sim$}}}}
\newcommand\gta{\mathrel{\hbox{\raise 0.6 ex \hbox{$>$}\kern
                   -1.7 ex\lower .5 ex\hbox{$\sim$}}}}

\newcommand{\ocen}{\mbox{$\omega$ Cen}}
\newcommand{\mzams}{\ensuremath{M_{\rm ZAMS}}}
\newcommand{\mem}[1]{\ensuremath{\mathrm{ #1}}}
\newcommand{\natlog}[2]{\ensuremath{#1\times 10^{#2}}} % a*10^b                                     
\newcommand{\msun}{\ensuremath{\, {\rm M}_\odot}}
\newcommand{\dex}{\ensuremath{\, \mathrm{dex}}}
\newcommand{\tab}[1]{Table\,\ref{#1}}
\newcommand{\mgvi}{\ensuremath{^{24}\mem{Mg}}}
\newcommand{\cdr}{\ensuremath{^{13}\mem{C}}}
\newcommand{\n}{\ensuremath{\mem{n}}}
\newcommand{\ose}{\ensuremath{^{16}\mem{O}}}
\newcommand{\czw}{\ensuremath{^{12}\mem{C}}}
\newcommand{\mgfu}{\ensuremath{^{25}\mem{Mg}}}

\newcommand{\nezwa}{\ensuremath{^{20}\mem{Ne}}}

\newcommand{\p}{\ensuremath{\mem{p}}}
\newcommand{\hevi}{\ensuremath{^{4}\mem{He}}}
\newcommand{\nvi}{\ensuremath{^{14}\mem{N}}}
\newcommand{\mdot}{\ensuremath{\dot{M}}}
\newcommand{\jahre}{\ensuremath{\, \mathrm{yr}}}
\newcommand{\mh}{\ensuremath{M_{\rm H}}}        % Masse der Wasserstoffschicht auf WD's             
\newcommand{\abb}[1]{Fig.\,\ref{#1}}
\newcommand{\kap}[1]{Sect.\,\ref{#1}}
\newcommand{\amlt}{\ensuremath{\alpha_\mathrm{MLT}}}
\newcommand{\mhe}{\ensuremath{M_{\rm He}}}
\newcommand{\ipr}{\mbox{$i$-process}}
\newcommand{\hei}{\ensuremath{^{1}\mem{H}}}
\newcommand{\hzw}{\ensuremath{^{2}\mem{H}}}
\newcommand{\hedr}{\ensuremath{^{3}\mem{He}}}
\newcommand{\lisi}{\ensuremath{^{7}\mem{Li}}}
\newcommand{\nfu}{\ensuremath{^{15}\mem{N}}}
\newcommand{\osi}{\ensuremath{^{17}\mem{O}}}
\newcommand{\oac}{\ensuremath{^{18}\mem{O}}}
\newcommand{\fne}{\ensuremath{^{19}\mem{F}}}
\newcommand{\neei}{\ensuremath{^{21}\mem{Ne}}}
\newcommand{\nezw}{\ensuremath{^{22}\mem{Ne}}}
\newcommand{\nadr}{\ensuremath{^{23}\mem{Na}}}
\newcommand{\lumh}{\ensuremath{L_{\rm H}}}

\newcommand{\lumhe}{\ensuremath{L_{\rm He}}}
\newcommand{\spr}{\mbox{$s$-process}}
\newcommand{\sprn}{\mbox{$s$ process}}
\newcommand{\apleq}{\ensuremath{\stackrel{<}{_\sim}}}
\newcommand{\apgeq}{\ensuremath{\stackrel{>}{_\sim}}}
 \newcommand{\iprn}{\mbox{$i$ process}}
\newcommand{\etal}{et~al.}
\slugcomment{Accepted for publication in ApJ: August 5, 2012}

\usepackage{natbib}

\shorttitle{CMD, (S)AGB stars and chemical evolution of \ocen}
\shortauthors{Herwig, VandenBerg, etal.}

\begin{document}

\title{From the CMD of $\omega$ Centauri  and (super-)AGB stellar models to
  a  Galactic plane passage gas purging chemical evolution scenario} 

\author{Falk Herwig\altaffilmark{1,2},
Don A.~VandenBerg\altaffilmark{1},
Julio F. Navarro\altaffilmark{1},
Jason Ferguson\altaffilmark{3},
and Bill Paxton\altaffilmark{4}
}
\altaffiltext{1}{Department of Physics \& Astronomy, University of Victoria,
       P.O.~Bos 3055, Victoria, B.C., V8W~3P6, Canada, 
       \email{fherwig@uvic.ca,vandenbe@uvic.ca}}
\altaffiltext{2}{Turbulence in Stellar Astrophysics Program, New Mexico
  Consortium, Los Alamos, NM 87544, USA}
\altaffiltext{3}{Department of Physics, Wichita State University
       Wichita, Kansas, USA \email{jason.ferguson@wichita.edu}}
\altaffiltext{4}{KITP/UC Santa Barbara, Santa Barbara,
  California, USA \email{paxton@kitp.ucsb.edu}}

\begin{abstract}
  We have investigated the color-magnitude diagram of $\omega$
  Centauri and find that the blue main sequence (bMS) can be
  reproduced only by models that have a of helium abundance in the
  range $Y=0.35$--$0.40$.  To explain the faint subgiant branch of the
  reddest stars (``MS-a/RG-a" sequence), isochrones for the observed
  metallicity ([Fe/H] $\approx -0.7$) appear to require both a high
  age ($\sim 13$ Gyr) and enhanced CNO abundances ([CNO/Fe] $\approx
  0.9$). $Y \approx 0.35$ must also be assumed in order to counteract
  the effects of high CNO on turnoff colors, and thereby to obtain a
  good fit to the relatively blue turnoff of this stellar
  population. This suggest a short chemical evolution period of time
  ($< 1$ Gyr) for \ocen. Our intermediate-mass (super-)AGB models are
  able to reproduce the high helium abundances, along with [N/Fe]
  $\sim 2$ and substantial O depletions if uncertainties in the
  treatment of convection are fully taken into account. These
  abundance features distinguish the bMS stars from the dominant
  [Fe/H] $\approx -1.7$ population.  The most massive super-AGB
  stellar models ($\mzams\geq6.8\msun$, $M_\mathrm{He,core} \geq
  1.245\msun$) predict too large N-enhancements, which limits their
  role in contributing to the extreme populations. In order to address
  the observed central concentration of stars with He-rich abundance
  we show here quantitatively that highly He- and N-enriched AGB ejecta
  have particularly efficient cooling properties.  Based on these
  results and on the reconstruction of the orbit of \ocen\ with
  respect to the Milky Way we propose the galactic plane passage gas
  purging scenario for the chemical evolution of this cluster. The bMS
  population formed shortly after the purging of most of the cluster
  gas as a result of the passage of \ocen\ through the Galactic disk
  (which occurs today every $\sim 40\mem{Myrs}$ for \ocen) when the
  initial-mass function of the dominant population had ``burned"
  through most of the Type II supernova mass range.  AGB stars would
  eject most of their masses into the gas-depleted cluster through
  low-velocity winds that sink to the cluster core due to their
  favorable cooling properties and form the bMS population. In our
  discussion we follow our model through four passage events, which
  could explain not only some key properties of the bMS, but also of
  the MS-a/RGB-a and the $s$-enriched stars.
\end{abstract}


\section{Introduction}
\label{sec:intro} 
Omega Centauri provides an especially valuable constraint on our
understanding of chemical evolution because of its unusual properties
and its proximity, which enables us to study its component stellar
populations in great detail over a very extended range in luminosity.
While it has been known for some time that the giants in this system
encompass a range in [Fe/H] from $\sim -2.2$ to $-0.5$
\citep[e.g.][]{brown:93,suntzeff:96}, the extensive spectroscopic
surveys carried out in the past decade, in particular \citep[][and
references
therein]{smith:00,hilker:04,kayser:06,vanloon:07,johnson:08,johnson:10,marino:12},
have established that the metallicity distribution rises sharply from
[Fe/H] $\approx -2.2$ to a strong peak at [Fe/H] $\approx -1.7$, with
a long tail that drops off to higher metal abundances containing
secondary peaks at [Fe/H] $\approx -1.4$, $-1.1$, and $-0.7$.  Type Ia
supernovae appear to have contributed to the chemical makeup of only
the most metal-rich stars given that their measured [$\alpha$/Fe]
abundance ratios are significantly reduced from the constant value of
$\approx 0.35$ found in stars having [Fe/H] $< -0.8$
\citep{pancino:02,origlia:03}.  The elevated $\alpha$-element
abundances over most of the range in metallicity (also see Kayser et
al.) implies that Type II supernovae were the major producers of these
elements.  Moreover, the rapid rise in the [La/Fe] ratio between the
most metal-deficient and the [Fe/H] $> -1.5$ populations (by a factor
of 3 according to Johnson \& Pilachowski) indicates that
$s$-processing, and therefore, intermediate-mass asymptotic-giant
branch (AGB) stars contributed significantly to the chemistry of
$\omega$ Cen after the dominant [Fe/H] $= -1.7$ population had formed.

Accompanying these spectroscopic advances have been equally impressive
improvements in the photometric data.  The detailed, and deep, color-magnitude
diagrams (CMDs) derived by \citet{lee:99,hughes:00,rey:04,bedin:04,sollima:05a,sollima:07,villanova:07,calamida:09}, and \citet{bellini:09,bellini:10}, among
others, have also established the existence of several discrete stellar
populations in $\omega$ Cen.  The most baffling one of them is a so-called
``blue main sequence" (bMS), discovered by \citep{anderson:97}, that is clearly
separated from, and bluer at a given magnitude than, the MS associated with
the dominant [Fe/H] $\approx -1.7$ population (see the CMDs reported by, e.g., 
Bedin et al.~and Villanova et al.).   What has made this discovery so puzzling
is the determination by \citet{piotto:05} that the former is more metal rich
than the latter by $\approx 0.3$ dex.  As discussed by Piotto et al., and
anticipated by Bedin et al.~ and \citet{norris:04}, the most obvious (only?) way
to reconcile these observations with stellar evolutionary theory is to infer
that the bMS stars have unusually high helium abundances ($Y > 0.35$).

Although it has not yet been possible to make a definitive connection
of the bMS through the subgiant region of the CMD to the RGB
\citep[but see][]{king:12}, \citet{johnson:10} have found that the
giants in their sample with ``extreme" abundances (i.e., those with
low C and O, together with high N, Na, and Al; see their Fig.~23)
constitute a similar proportion of all giants ($\approx 27\%$) as the
fraction of all MS stars that are located on the bMS.  Furthermore,
since (for the most part) their radial distributions are quite
similar, it is tempting to conclude that the aforementioned giants are
the descendants of stars that had occupied the bMS earlier in their
evolutionary history.  However, because the O-poor stars do not
exhibit a correlation with [La/Fe], while the O-rich stars do, and
because there is no apparent correlation of the La abundance with
radius, in contrast to the observed trend for the O-poor stars,
Johnson \& Pilachowski suggest that the observed light element
abundances reflect both (i) the retention by $\omega$ Cen of the
ejecta of AGB stars and (ii) {\it in situ} mixing on the RGB.

There is a second feature of the CMD that is potentially quite
challenging to explain, and that is the extension of the reddest giant
branch (independently discovered by \citet{lee:99} and by
\citet{pancino:00}, who gave it the name ``RGB-a" that has since been
used to refer to it), through the subgiant branch \citep{ferraro:04}
to very faint magnitudes on the MS \citep{bellini:10}.  Because this
is is the most metal-rich of the discrete populations that have been
identified, it is presumably also the youngest one.  Consequently, its
age provides a key constraint on the timescale over which most of the
chemical evolution took place in $\omega$ Cen.  Because the turnoff of
this fiducial sequence is so faint, the age of this stellar population
must be quite old, implying that the bulk of the star formation
occurred over a rather short interval of time.  Indeed, the reason why
some investigations \citep[e.g.][]{hilker:04,rey:04} derived an
extended star formation history ($\sim 3$--4 Gyr) in their analyses is
that they had mistakenly adopted a bright turnoff for the RGB-a
population.  However, it remains unclear whether the different stellar
populations of \ocen\ have the same age to within a few $\times 10^8$
yr \citep{D'Antona2011,valcarce:11}, or they span a range in age of
$\lta 2$ Gyr \citep{sollima:05a} or as much as 3--5 Gyr
\citep{stanford:06,villanova:07}.

Among stellar evolution models of different mass ranges, intermediate-mass and
super-AGB stars have been proposed as the source of the He-enriched
and otherwise extreme abundance patterns associated with the bMS.
However, models of these types of stars do not always
reproduce the required high He, high N, and low O abundance associated
with the bMS in a quantitative way. For example, the intermediate-mass AGB
models by \citet[][specifically the $Z=\natlog{6}{-4}$ case
that is applicable to the first, most metal-deficient generation in
\ocen]{ventura:09} predict $Y=0.329$ and
[N/Fe]$=1.84$ for $\mzams=5.0\,\msun$ and $Y=0.360$ and [N/Fe]$=1.36$ for
$\mzams=6.0\,\msun$. The super-AGB models by \citet{ventura:11} produce a
maximum ejecta He abundance of $Y=0.36$, but information for N is not
available.  The envelope He abundance at the end of the second
dredge-up in the super-AGB models by \citet{siess:07} reach $0.37$ for
non-overshooting models, and $0.38$ for models with core overshooting.
\citet{siess:10} do not provide a model for $Z=\natlog{6}{-4}$, but
interpolating between their super-AGB thermal pulse calculations
for $Z=0.001$ and $Z=10^{-4}$ yield He mass fractions in the
ejecta of $Y\sim 0.33$, while N is enhanced by a factor of $10$ to $100$.
Moreover, those models with the largest increases in the nitrogen
abundance also have the smallest O depletion, and models with O
depletions of $\sim 1 \dex$ have N enhancements of only $\sim
1 \dex$.  While existing models clearly point qualitatively in the
right direction, it is not clear to us what it would take to reproduce
the ``extreme'' abundance mixtures that are associated with the bMS in \ocen\
in a quantitative sense. 

Producing He-abundances in intermediate-mass AGB ejecta that
exceeds the He-abundance of the ``extreme'' abundance mix in any
significant way appears to be very difficult within model
calculations. This does not leave a lot of room (if any) for dilution of
the AGB ejecta before the formation of the blue main-sequence
stars. 
Dilution has been suggested, however, in order to reproduce the abundance
anti-correlations \citep{d'ercole:08}. We are considering here
rather the origin and evolution of one sub-population at a time. The
presentation of Na-O abundances by sub-populations identified with a
four-criterion cluster analysis by
\citet[][Fig.\,6]{Gratton:2011kr} rather suggests that the
anti-correlation is the result of the superposition of discrete Na and
O abundance in different sub-populations, and as we will argue later
(\kap{sec:final-ONa}) dilution may not be needed. 
We therefore investigate the evolution of first-generation
intermediate-mass AGB and super-AGB stars with the goal to generate the ``extreme''
abundance mix without taking into account any dilution.

In any case, the AGB ejecta may preferentially converge in the cluster
center via AGB cooling flows \citep[][and
refs.\ there]{DAntona:2011dm}. Such gas may have to be isolated, for
example through supernova-induced clearing of intracluster gas that is
ejected from stars having initial masses outside the range that
encompasses intermediate-mass AGB and super-AGB stars
\citep{Charlie2011a}, while the tidal stripping of old stars increases
the ratio of second- to first-generation stars
\citep{bekki:11}. However, complete SN gas purging makes it difficult
to envisage how a spread in [Fe/H], like that observed in \ocen, is
produced.

The present study has been undertaken to address some of the issues
described above.  In \S 2, newly computed sets of isochrones and
zero-age horizontal branch (ZAHB) loci for different values of $Y$,
$Z$, and heavy-element mixtures are compared with the CMD of $\omega$
Cen in order to illustrate how the interpretation of the observations
would be affected by the assumed variations in the chemical abundances,
and to make some assessment of the age and helium content of the stars
that belong to the bMS and MS-a components.  \S 3 investigates under which
assumptions the predicted yields from $> 5\msun$ AGB stars can
be made to agree with the observed/inferred high $Y$, high-N, low-O abundances.
Finally, a short summary of the main results of this study is given
in \S 4, which also discusses in some detail how our stellar evolutionary
results together with the efficient cooling properties of the ejecta from
intermediate-mass AGB stars may naturally explain the formation of additional
populations like the bMS.  A key point in our proposed scenario is the
periodic purging of the gas from globular clusters (or dwarf galaxies)
that is expected to occur throughout their evolutionary histories whenever
their orbits cause them to pass through the Galactic
plane. Implications of this scenario are briefly summarized in \S 5.

\section{Color-Magnitude Diagram Considerations}
\label{sec:cmd}

Several grids of stellar evolutionary tracks were computed using a
significantly updated version of the Victoria code \citep{vandenberg:12}
which now treats the diffusion of helium (but not the metals) as well as
extra mixing below envelope convection zones (when they exist) using methods
very similar to those described by \citet{proffitt:91}.  All of the
model computations reported in this study assumed a value of
$\alpha_{\rm MLT} = 2.007$ for the usual mixing-length parameter, so
as to satisfy the solar constraint when the mix of heavy elements
derived for the Sun by \citet{asplund:09} is assumed.  The
abundances of the individual $\alpha$-elements in this mixture were
increased by approximately the amounts found in very metal-deficient
stars according to \citet{cayrel:04}, resulting in the ``standard"
metals mix (see the second column of Table~\ref{tab:abundances_pops})
that applies to normal Population II stars, including (presumably) the
dominant [Fe/H] $\approx -1.7$ component of $\omega$ Cen.
\begin{deluxetable}{ccccc}
\tabletypesize{\footnotesize}
\tablecaption{Chemical Abundance Ratios \label{tab:abundances_pops}}
\tablewidth{0pt}
\tablehead{
 \colhead{}    &  \multicolumn{4}{c}{[m/Fe]} \\
 m & Standard &  Extreme & $\omega$ Cen\tablenotemark{a}
 & NGC$\,$2808\tablenotemark{b} 
} 
\startdata
\phantom{e}C & \phantom{$+$}0.0 & $-0.7$ & $-0.5$ & $-0.7$ \\
\phantom{e}N & \phantom{$+$}0.0 & $+2.0$ & $+1.5$ & $+2.0$ \\
\phantom{e}O & $+0.5$         & $-0.5$ & $-0.5$ &        \\
Ne           & $+0.3$ & \phantom{$+$}0.0 &        &        \\
Na           & \phantom{$+$}0.0 & $+0.8$ & $+0.6$ & $+0.8$ \\
Mg           & $+0.3$ & \phantom{$+$}0.0 &        & $+0.1$ \\
Al           & \phantom{$+$}0.0 & $+1.1$ & $+1.1$ & $+1.1$ \\
Si           & $+0.4$         & $+0.4$ & $+0.4$ &        \\
\phantom{e}P & \phantom{$+$}0.0 & \phantom{$+$}0.0 &  &      \\
\phantom{e}S & $+0.3$         & $+0.3$ &        &        \\
Cl           & \phantom{$+$}0.0 & \phantom{$+$}0.0 & &       \\
Ar           & $+0.3$         & $+0.3$ &        &        \\
\phantom{e}K & \phantom{$+$}0.0 & \phantom{$+$}0.0 &  &      \\
Ca           & $+0.3$         & $+0.3$ & $+0.3$ & $+0.3$ \\
Ti           & $+0.3$         & $+0.3$ & $+0.3$ &        \\
Cr           & \phantom{$+$}0.0 & \phantom{$+$}0.0 &  &      \\
Mn           & \phantom{$+$}0.0 & \phantom{$+$}0.0 &  &      \\
Ni           & \phantom{$+$}0.0 & \phantom{$+$}0.0 &  &      \\
\enddata
\tablenotetext{a}{For O-poor giants \citep{johnson:10}.}
\tablenotetext{b}{For a blue MS star \citep{bragaglia:10}.}
\end{deluxetable}

The third column of Table~\ref{tab:abundances_pops} lists the
``extreme" heavy-element mixture that has been assumed in some of the
comparisons of observations with models to be discussed shortly.
Remarkably, multiple main sequences have also been discovered in
NGC$\,$2808 \citet{Piotto:07}, and the chemical abundances derived by
\citet{bragaglia:10} for one of its bluest MS stars appear to be quite
similar to those found in what are believed to be the red-giant
descendants of bMS stars in $\omega$ Cen (compare the entries in the
4th and 5th columns).  To maximize the impact of these abundance
anomalies on the predicted properties of stellar models, we have
chosen to adopt the largest of the [$m$/Fe] values (in an absolute
sense) that have been determined for $\omega$ Cen and NGC$\,$2808 in
our ``extreme" heavy-element mix (column 3).  However, we note that,
although \citet{marino:11} have found that the [O/Fe] value varies
from $-0.6$ to $+0.9$ at metallicities appropriate to bMS stars, the
[CNO/Fe] ratio spans the relatively small range of $\approx +0.4$--$0.7$
\citep[see][]{marino:12}.  This suggests that the CNO abundances
listed in the third column of Table 1 are somewhat more extreme than
those found in \ocen\ stars having [Fe/H] $\approx -1.5$.  On the 
other hand, \citet[][see their Fig.~3]{marino:12} find that the
[CNO/Fe] value ranges between $+0.7$--$0.9$ at [Fe/H] $\sim -1.0$, and
as discussed below, such high values appear to be necessary to explain
the faint subgiant branch of the MS-a/RG-a population.

  Also worth mentioning is the fact that opacity data for both mixtures
were generated using the code described by \citet{ferguson:05} for the
low-temperature regime, while complemetary high-temperature tables
\citep[similar to those reported by][]{iglesias:96} were obtained via
the OPAL website\footnote{See http://opalopacity.llnl.gov}.  Note, as
well, that the interpolation program developed by P.~Bergbusch, which
is similar to that described by \citet{vandenberg:06}, but with the
significant improvements \citep[see][]{vandenberg:12}, was used to produce
all of the isochrones considered in this investigation.

\begin{figure}
%\plotone{Figures/ocenf1.pdf}
   \includegraphics[width=0.48\textwidth]{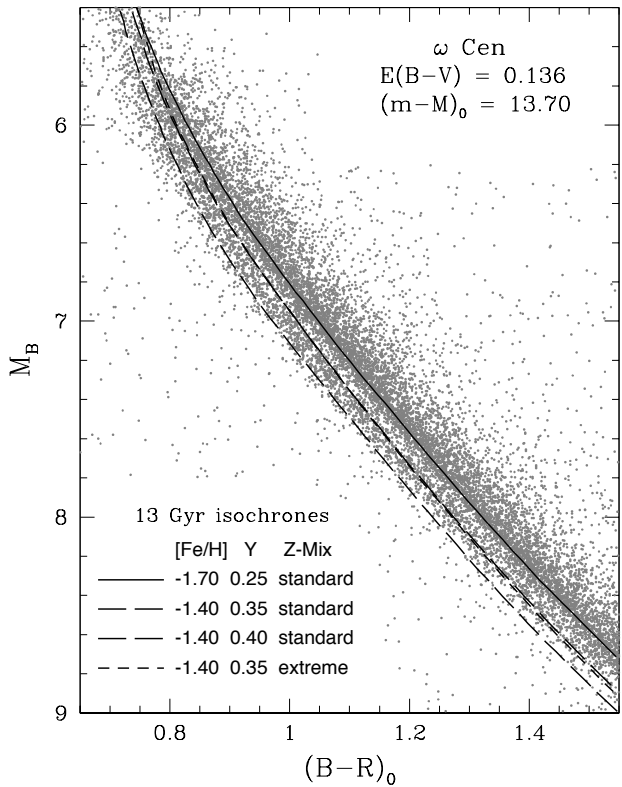}
%\plotone{fig1.pdf}
\caption{Comparison of the lower main-sequence extensions of isochrones for the
indicated age and chemical abundances with the photometry of $\omega$ Cen
reported by \citet{sollima:07}.  The adopted values of $E(B-V)$ and the true
distance modulus are noted in the upper right-hand corner.}
\label{fig:fig1}
\end{figure}

To begin our analysis, we first compare, in \abb{fig:fig1}, the
main-sequence segments of several isochrones with the lower MS
photometry of $\omega$ Cen given by \citet{sollima:07}, based on VLT
FORS1 observations.  The latter have been converted to the
[$(B-R)_0,\,M_B$]-plane assuming $(m-M)_0 = 13.70$
\citep{bellazzini:04}, $E(B-V) = 0.136$ \citep{schlegel:98}, and
$A_B = 4.07\,E(B-V)$, $A_R = 2.44\,E(B-V)$ \citep{mccall:04}.  Bellazzini
et al.~have argued that the \citet{lub:02} determination of $E(B-V) = 0.11
\pm 0.01$ mag is the current best estimate of the foreground
reddening, and that may well be the case.  However, the main purpose
of Fig.~1 is to determine the value of $Y$ that is needed to reproduce
the location of the bMS stars {\it relative} to the main sequence of
the dominant [Fe/H] $\approx -1.7$ population.  One could obviously
adopt the lower value of $E(B-V)$ and then increase the predicted
$B-R$ colors by 0.026 mag to obtain an identical fit of the models to
the observations in a differential sense.\footnote{To transpose the
  models from the theoretical to the observed plane, we have used the
  color--$T_{\rm eff}$ relations derived by L.~Casagrande
  \citep[see][]{vandenberg:10} from synthetic spectra
  based on the latest MARCS model atmospheres \citep{gustafsson:08}.}
In fact, fainter than $M_B = 6.5$, the synthetic colors had to be
adjusted by $\delta(B-R) = 0.02(M_B - 6.5)$ in order for the [Fe/H] $=
-1.70$, $Y = 0.25$ isochrone (the solid curve) to accurately reproduce
the observed MS over the entire magnitude range that has been ploteed.
(Such a color offset could arise if, among other possibilities, the MARCS
model atmospheres predict too much flux at short wavelengths or there are
systematic errors in the model $T_{\rm eff}$ scale due to the treatment
of convection or the surface boundary conditions.)

When such a color correction is applied (to all of the model loci),
one finds that the bMS stars in $\omega$ Cen are bracketed by
isochrones for $Y = 0.35$ and $Y = 0.40$ (the long-dashed curves) on
the assumption of the observed [Fe/H] value \citep[$-1.4$,
][]{piotto:05} and the ``standard" Pop.~II metals mixture. 
(We note that $Y=0.39\pm0.02$ has been obtained by \citep{king:12}
from a fit of isochrones to their improved CMD.)
 Indeed,
these results are fully consistent with the findings of Piotto et
al.~(see their Fig.~7) and \citet[][their Fig.\,9]{sollima:07}, who
performed similar comparisons using completely independent theoretical
computations.  (While we have opted to plot 13 Gyr isochrones, mainly
because their extensions to brighter magnitudes are shown in a
subsequent figure, the location of the lower MS is obviously
independent of the assumed age.)  Interestingly, fainter than $M_B
\sim 5.7$, isochrones for the ``standard" and the ``extreme"
heavy-element mixtures are essentially identical when the same values
of $Y$ and [Fe/H] are assumed (compare the short-dashed curve with the
long-dashed curve for the same helium abundance).  \citep[Of course,
differences would be evident if photometric filters are used that are
sensitive to the abundances of CNO and/or other heavy elements; see,
e.g.][]{bellini:10,sbordone:11,milone:12a}  As others, including those
mentioned above and e.g., \citet{norris:04}, have concluded, it does
not appear to be possible to explain the location of the bMS stars in
$\omega$ Cen (and NGC$\,$2808) without invoking high $Y$ --- provided
that, indeed, chemical abundance differences are primarily responsible
for the CMD anomalies.  Indeed, we explored the impact of varying the
abundances of O, Mg, and Si using isochrones presented by
\citet{vandenberg:12}, which allow for variations in the abundances of
several metals in turn, but we were unable to find a satisfactory
alternative explanation (i.e., other than high $Y$) for the location
of the bMS relative to the main-sequence fiducial of the dominant
[Fe/H] $\sim -1.7$ population.

There is one aspect of Fig.~\ref{fig:fig1} that deserves further comment.  The 
bMS stars are clearly separated from the dominant red MS at $6.7 < M_B < 7.7$,
but not at lower luminosities, where the two populations appear to merge.  By
contrast, the isochrones remain parallel to one another over the entire range
in $M_B$ that is considered.  One can speculate that this may be a
color--$T_{\rm eff}$ relations effect; i.e., that at sufficiently cool
temperatures, stars having the ``extreme" metals mixture are redder at a fixed
$T_{\rm eff}$ and gravity than stars having normal Pop.~II abundances.  To
investigate this possibility, it would be necessary to compute proper model
atmospheres and synthetic spectra for the two heavy-element mixtures and then
compute the fluxes in the various filter passbands.  Additional work along
these lines would certainly be worthwhile.

\begin{figure} 
%  \plotone{Figures/ocenf2.pdf}
%  \plotone{fig2.pdf}
   \includegraphics[width=0.48\textwidth]{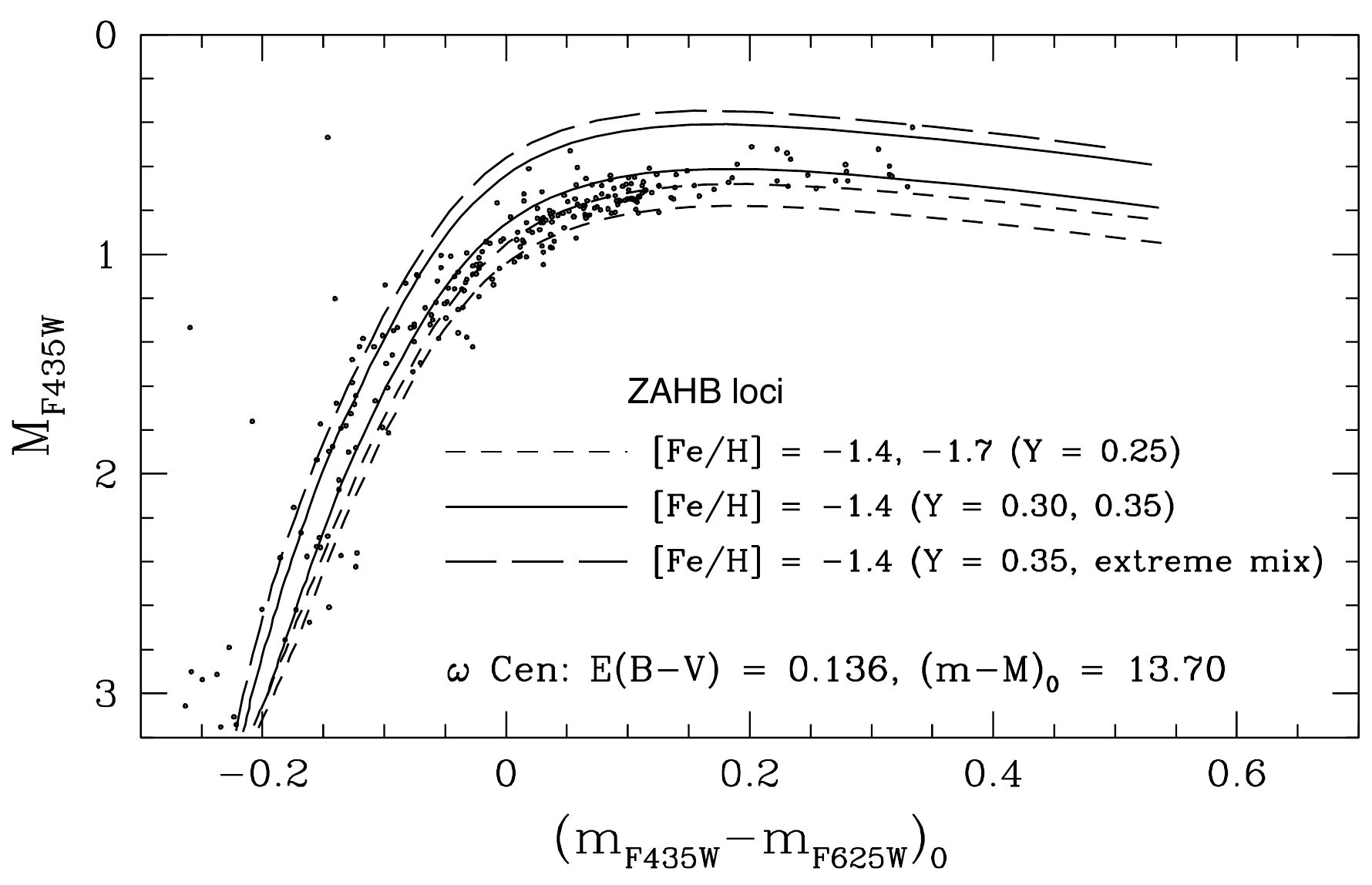}
  \caption{Comparison of several ZAHB loci for the indicated
    properties with the horizontal-branch stars in $\omega$ Cen having
    $M_B < 3.2$ \citep[from][]{bellini:09} on the assumption of the
    reddening and distance modulus that were assumed in the previous
    figure, as noted. As mentioned in the text, the predicted colors
    had to be increased by $+0.04$ mag in order for the models to
    reproduce the HB population in the blue tail satisfactorily.}
\label{fig:fig2}
\end{figure}

Before turning to an examination of the upper-MS to lower-RGB stars in $\omega$
Cen, it is instructive to examine its HB population.  \abb{fig:fig2}
compares the HB stars that have $M_{\rm F435W} < 3.2$ from the published  
\citet{bellini:09} {\it HST} data set with ZAHB loci that have been
calculated for indicated values of [Fe/H] and $Y$, using the numerical methods
described by \citet{vandenberg:00}.  As indicated, the same reddening and
distance modulus that were assumed in the previous figure have been adopted
here, along with $A_{\rm F435W} = 4.081\,E(B-V)$ and $A_{\rm F625W} =
2.637\,E(B-V)$ from \citet{sirianni:05}.  However, besides the reddening
adjustment, the predicted colors had to be corrected by $+0.04$ mag in order to
achieve a satisfactory fit of the ZAHB loci to the steeply sloped blue stars.
It is not clear why an additional zero-point correction is needed, as the same
ZAHBs appear to provide an excellent match to the $F606W, F814W$ observations
of \citet{anderson:08} without having to apply any {\it ad hoc} color
shift\footnote{According to A.~Sollima (2010, private communication), the HB
models that were compared with an eariler reduction of the same observations in
the Sollima et al.~(2005) study also required some adjustment of the predicted
colors.}.  Be that as it may, this concern is not of particular importance for
the present analysis, given that we are primarily interested in the comparison
of predicted and observed CMDs in a differential sense.  Note that, to
accomplish the transformation of the models to the observed plane, we have used
the color--$T_{\rm eff}$ relations applicable to ACS photometry that were
presented by \citet{bedin:05}, and kindly provided to us by S.~Cassisi (2011,
priv.~communication).

As already shown by \citet{sollima:05b}, most of the HB stars in
$\omega$ Cen having $(m_{\rm F435W}-m_{\rm F625W} > 0.0$ appear to be
matched quite well by ZAHB loci for [Fe/H] values from $\sim -1.7$ to
$\sim -1.4$ and $Y \approx 0.25$ (the short-dashed curves).  Although
some of the observed stars lie between the two solid curves, which
represent ZAHBs for [Fe/H] $= -1.4$ and $Y = 0.30, 0.35$, it is not
obvious whether they truly have higher helium abundances or they have
simply evolved to their current CMD locations from initial structures
on ZAHBs for lower values of $Y$.  The most noteworthy result of
Fig.~\ref{fig:fig2} is the lack of any stars brighter or bluer than
the ZAHBs for $Y = 0.35$ and [Fe/H] $= -1.4$ --- assuming either the
``standard" or the ``extreme" metals mixtures (i.e., the brightest of
the solid curves and the long-dashed curve, respectively).  ZAHB loci
for lower [Fe/H] values and/or higher helium abundances would be even
brighter at a given color than the latter.  Thus, if $\omega$ Cen does
contain stars having $Y > 0.35$, they must evolve to ZAHB locations
well down the blue tail.  This would perhaps not be too surprising
given that, at the same age and [Fe/H] value, the turnoff mass is
predicted to be $\approx 0.12$ solar masses less for stars having $Y =
0.35$ than those for $Y = 0.25$, nearly independently of the metals mixture that
is assumed.  If similar, and significant, amounts of mass loss occur
on the RGB, then stars having high $Y$ {\it should} evolve into ZAHB
structures that are much hotter than those for normal helium
abundances.  This has already been appreciated by a number of others
\citep[e.g.][]{busso:07,d'antona:08}, and in fact, \citet{cassis:09}
have suggested that the densely populated clump of HB stars at
$m_{F435W} \sim 18.6$ ($M_{F435W} \sim 4.4$) in the \ocen\ CMD may be
He-rich stars.

\abb{fig:fig3} makes a number of interesting comparisons of theoretical
isochrones with the \citet{bellini:09} ACS data for upper-MS, subgiant, and
lower-RGB stars of $\omega$ Cen.  (Only a fraction of the total number of
observed stars have been plotted, for the sake of improved clarity.)  Curiously,
it was not necessary to apply any correction to the isochrone colors, as derived
from the transformations by \citet{bedin:05}, in order to obtain quite a good
match to the observed MS stars, which leads us to wonder whether the difficulty
noted above concerning the colors of ZAHB models should be attributed to a small
problem with these color--$T_{\rm eff}$ relations (for just warm stars) or
whether our interpretation of the observed HB is incorrect.  The former
explanation would seem to be the most likely one given that, if the predicted
colors are not adjusted to the red, the core He-burning stars of $\omega$ Cen
would lie below, or redward, of all of the ZAHB loci (assuming that the adopted
distance modulus is accurate), which would be very difficult to explain.
\begin{figure}
%\plotone{Figures/ocenf3.pdf}
   \includegraphics[width=0.48\textwidth]{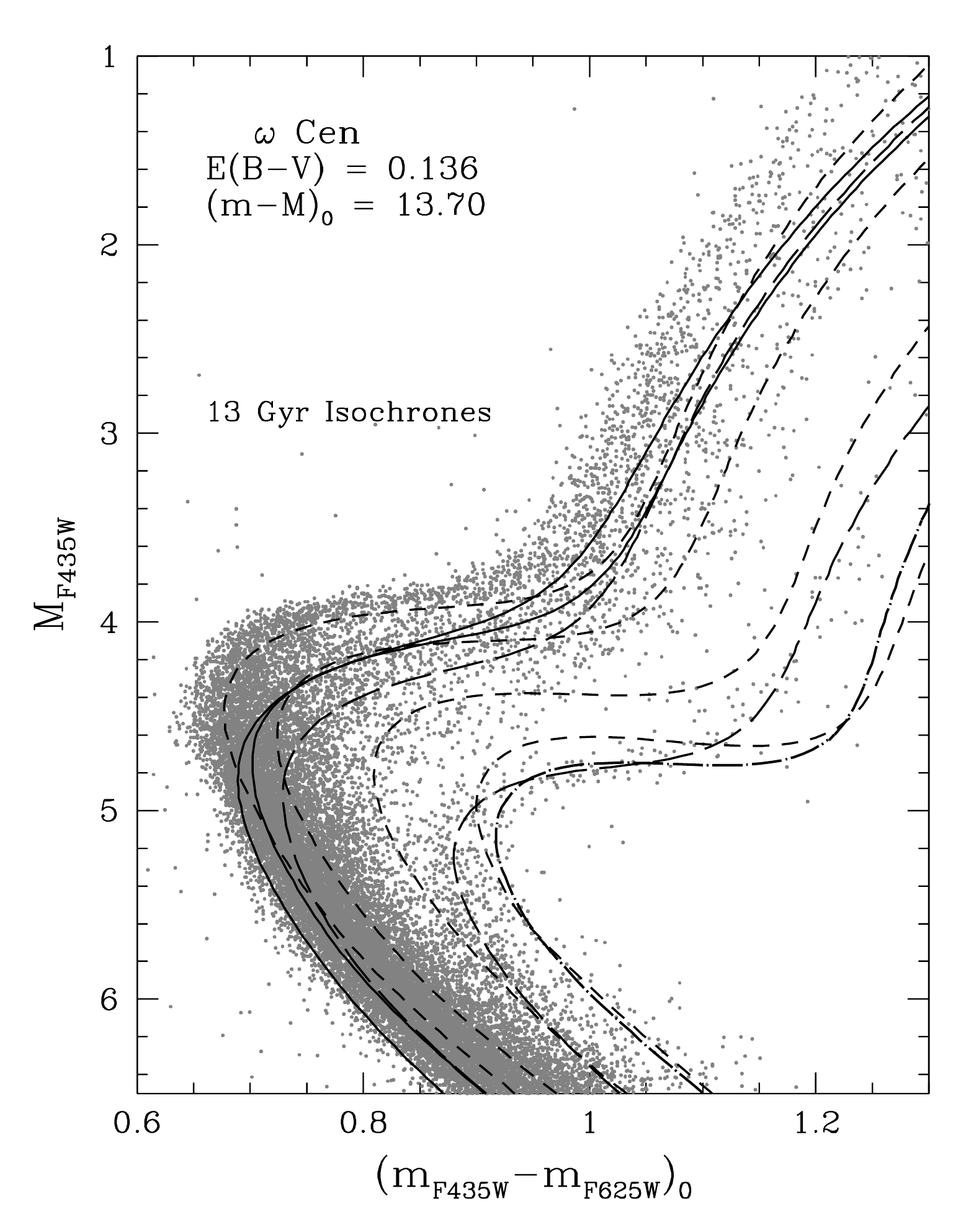}
%\plotone{fig3.pdf}
\caption{Comparison of 13 Gyr isochrones with the upper main-sequence, subgiant,
and lower red-giant-branch populations of $\omega$ Cen.  The short-dashed
curves assume $Y = 0.25$, the ``standard" mix of heavy elements (see 
Table~\ref{tab:abundances_pops}), and [Fe/H] $= -1.7$, $-1.4$, $-1.0$, and $-0.7$, in the
direction from left to right.  The solid curves were computed for [Fe/H]
$=-1.4$ and the same metals mix, but for $Y = 0.35$ and 0.40: the isochrone
for the higher helium abundance has a bluer turnoff.  The long-dashed
loci represent isochrones for $Y = 0.35$, the ``extreme" mix, and [Fe/H]
$= -1.4$ and $-0.7$, in the direction from left to right.  The dot-long-dashed
isochrone assumes the ``extreme" mixture, as well, but a normal helium
abundance ($Y = 0.250$).}
\label{fig:fig3}
\end{figure}

The short-dashed curves represent 13 Gyr isochrones for $Y = 0.25$, and
[Fe/H] $ = -1.7$, $-1.4$, $-1.0$, and $-0.7$ (in the direction from left to
right), on the assumption of the ``standard" metals mix.  (An isochrone for
the lowest metallicity and an age of 12.5 Gyr would actually provide the closest
match to the densest distribution of stars which define the brightest subgiant
branch (SGB), and is therefore our best estimate of the age of the most
metal-poor population of $\omega$ Cen.)  The RGB segments of these isochrones
appear to be too red, by $\sim 0.05$ mag, which suggests that either the
model $T_{\rm eff}$ scale is somewhat too cool and/or the adopted color
transformations for low-gravity stars are slightly too red.  However, {\it in a
relative sense}, the difference in the colors of the predicted RGBs for [Fe/H]
$= -1.7$ and $-1.0$, at a fixed color, is comparable to the observed width of
the main distribution of giant stars.

Isochrones represented by the solid curves and the bluest of the long-dashed
loci extend to higher luminosities the same isochrones that were plotted in
Fig.~\ref{fig:fig1}. As emphasized by \citet{sollima:05b}, among
others, variations in the helium content have only minor effects on
the luminosity and morphology of the SGB. This is demonstrated by the
fact that the two solid isochrones (for $Y = 0.35$ and 0.40) and
the short-dashed isochrone (for $Y = 0.25$) --- all of which were
computed for [Fe/H] $= -1.4$ on the assumption of the ``standard" mix
of heavy elements --- are nearly coincident along the subgiant branch.
On the other hand, the isochrones for $Y = 0.35$ and [Fe/H] $= -1.4$
that has been computed for the ``extreme" metals mix (the aforementioned
long-dashed isochrone) has a significantly fainter SGB at a
same age (due to these models having a much higher [CNO/Fe] abundance
ratio). 

Although \citet{king:12} have found that the SGB tentatively
identified as the extension of the of the bMS can be matched by
isochrones for high Y and a normal $\alpha$-enhanced mixture, the
sensitivity of subgiant luminosities, at a fixed age, to the total
abundance of the CNO elements \citep[e.g., see][]{cassisi:08}, clearly
complicates the determination of the ages of $\omega$ Cen SGB stars if
derived from their locations relative to theoretical isochrones for
different metallicities \citep[e.g.,][]{hilker:04, stanford:07}. In
order to have robust ages, even in a differential sense, it is of
critical importance to take into account any star-to-star variations
that exist in both [Fe/H] and [CNO/Fe].  According to
\citet{marino:12}, the mean [CNO/Fe] ratio varies from $\approx 0.25$
at [Fe/H] $= -1.9$ to $\approx 0.75$ at [Fe/H] $ = -1.0$ (which is
very difficult to understand; see their discussion of this point),
with a dispersion of about 0.2 to 0.4 dex at a given metallicity.
Such variations of [CNO/Fe] at a fixed iron abundance will affect
predicted ages at the level of $\sim 0.6$--1.3 Gyr
\citep{vandenberg:12}.

Finally, Fig.~\ref{fig:fig3} shows that the most metal-rich of the discrete
stellar populations in $\omega$ Cen has a fainter SGB, and an appreciably
bluer turnoff, than those of a 13 Gyr isochrone for $Y = 0.25$ and [Fe/H]
$= -0.7$ (``standard" mix).  Surprisingly, the reddest of the long-dashed
isochrones, for $Y = 0.35$ and [Fe/H] $= -0.7$ (``extreme" mixture), does
reproduce these observations quite well --- though the inferred value of $Y$
cannot be determined very precisely because the upper MS and turnoff of the
MS-a population is not very well defined in the photometry that we have
used.\footnote{The fiducial sequence for these stars {\it is} clearly defined
in the CMD reported by Bellini et al.~(2010), but those data are presented
for filter bandpasses that we are unable to model at the present time due to
the lack of suitable color transformations.  Importantly, Bellini et al.~find
a single, well-defined turnoff for the MS-a sequence, which justifies our
decision to fit isochrones to roughly the middle of this distribution of
stars.}  On the other hand, isochrones for the same metallicity and a normal
helium content ($Y = 0.25$) are clearly too red and they fail to reproduce
the slope of the subgiant branch (note the location of the dot-long-dashed
curve).  Isochrones for high $Y$ do not suffer from such difficulties.  Thus,
it appears to be a fairly robust stellar evolution prediction that the
MS-a/RG-a stars have high CNO {\it and} He abundances.  

These results support previous suggestions.  In particular,
\citet{marino:11} noted that ``as with bMS stars, the colors and the
metallicity of the MS-a are consistent with being populated by He-rich
stars \citep{norris:04,bellini:10}.  Moreover, based on observations
taken in many different ACS bandpasses, the aforementioned Bellini et
al.~paper suggested that this population has ``peculiar" CNO
abundances.  Indeed, the predicted age of the MS-a/RG-a stars will be
less than the age of the universe only if they have high CNO
abundances (assuming that our estimate of the \ocen\ distance is
accurate).  Furthermore, the slope of the subgiant branch, which is
well defined in Fig.~3, can be reproduced by the models only if high
$Y$ is also assumed.  As far as our contention that a high helium
abundances is needed to match the turnoff color is concerned, we note
that (i) the $T_{\rm eff}$ scale of our models agrees well with that
derived for solar neighborhood stars (having {\it Hipparcos}
parallaxes) over the entire range in [Fe/H] encompassed by them
\citep[see][]{vandenberg:10}, and (ii) isochrones for close to the
primordial abundance of helium and standard Pop.~II metal abundances
provide a good fit to the [Fe/H] $= -1.7$ population.  Stellar models
should be quite reliable in a differential sense.

Our CMD analysis suggests that the different stellar populations in \ocen\ have
close to the same age, and hence that the chemical evolution in this system
must have occurred over a relatively short period of time ($\lta 1$ Gyr).
(For the MS-a/RG-a stars to be appreciably younger than the universe, they
they would neeed to have a higher [Fe/H] value and/or higher CNO or helium
abundances than those assumed in the best-fit isochrone.)  Especially
intriguing is the very real possibility that the most metal-rich component
has a high helium abundance.  Is there, then, a connection between these stars
and the bMS?  Before offering some suggestions on how the chemical evolution
in $\omega$ Cen may have proceeded (in \S 4), we will first examine whether
the chemical yields from AGB models that form out of gas having the ``standard"
mix of heavy elements (column 2 of Table~\ref{tab:abundances_pops}) will be
similar to the ``extreme" mixture (col.~3).

\section{AGB and super-AGB stellar evolution with uncertain convective
mixing physics}
\label{sec:agb}
% /rpod2/fherwig/M/Set_OCen_standard
% m5=ms.star_log('M5.00Z5.9E-4_lowMdot/LOGS',clean_starlog=False)
% m5_mlt=ms.star_log('M5.00Z5.9E-4_MLT_lowMdot/LOGS',clean_starlog=False)

Our analysis of the CMD of $\omega$ Cen suggests that somehow out of
the first, and dominant, [Fe/H] $\approx -1.7$ population, in which the
metals have the ``standard" $\alpha$-enriched distribution, at least two
distinguishable new stellar populations have been generated. These are
the [Fe/H] $\approx -1.4$ bMS stars and their counterparts on the RGB,
which appear to have helium abundances between $Y=0.35$ and $0.40$, and
the so-called ``MS-a/RG-a" population at an even higher Fe abundance
($\approx -0.7$) that seems to have a similarly enhanced He abundance
as well as enhanced C$+$N$+$O, judging from the comparison of isochrones
with the observed photometry.

In this section we wish to address the question of whether current AGB
and/or super-AGB models are able to produce ejecta out of which the bMS
population, and possibly the MS-a population, can form.

\subsection{Model assumptions}
We have calculated several intermediate-mass AGB and super-AGB stellar
evolutionary sequences with an initial abundance distribution (see
\tab{tab:don_A09C}) that closely resembles the abundances found in
the dominant generation of stars at [Fe/H] $\approx -1.7$. The tabulated
mass-fraction abundances for the metals are, in fact, equivalent to the 
abundances that have been labelled as the ``standard" mix in
\tab{tab:abundances_pops}, and the adopted isotopic ratios are from
\citet{asplund:09}.

 We use the MESA stellar evolution code
\citep[rev.\,2941,][]{paxton:11}. The MESA paper by Paxton
et\,al.\ already contains verification cases for a wide range of stellar
evolution cases, including thermal pulses and dredge-up in AGB
stars. In addition, we have now also run massive AGB models with the
same initial abundances and similar enough physics assumptions as
those chosen for the grid of models including massive AGB stars by
\citet{herwig:04a} obtained with the EVOL stellar evolution code, and
we find the MESA results again to be in reasonable agreement.

We adopt a customized nuclear network
(identified as \texttt{sagb\_NeNa.net}), which is based on the MESA
network \texttt{agb.net}.  To be more specific, we consider 23
chemical species (including the major isotopes) for the elements from
H to Na, ending with \mgvi, as well as the $pp$-chain, the CNO cycles,
the NeNa cycle, He-burning, the $\cdr(\alpha,\n)\ose$ reaction,
and the $\czw + \mbox{\czw} \rightarrow \nezwa + \alpha$ reaction.
The nuclear reaction rates have been taken from the NACRE compilation
\citep{angulo:99}. 
\begin{deluxetable}{ll}
\tabletypesize{\footnotesize}
\tablecaption{Initial abundance distribution for the AGB stellar
  evolution calculations. 
\label{tab:don_A09C}}
\tablewidth{0pt}
\tablehead{ Isotope & Initial mass fraction\tablenotemark{a}    }
\startdata
\hei& $   0.7494 $\\
\hzw & $   1.4576\times 10^{-5} $\\
\hedr & $  2.3547\times 10^{-5} $\\
\hevi & $  0.2500 $\\
\lisi & $  4.6251\times 10^{-1} $\\
\czw & $   4.7392\times 10^{-5} $\\
\cdr & $   5.7061\times 10^{-7} $\\
\nvi & $   1.4014\times 10^{-5} $\\
\nfu & $   3.2165\times 10^{-8} $\\
\ose & $   3.6666\times 10^{-4} $\\
\osi & $   1.3930\times 10^{-7} $\\
\oac & $   7.3508\times 10^{-7} $\\
\fne & $   2.5000\times 10^{-9} $\\
\nezw & $  4.6817\times 10^{-5} $\\
\neei & $  1.1223\times 10^{-7} $\\
\nezw & $  3.4425\times 10^{-6} $\\
\nadr & $  5.9273\times 10^{-7} $\\
\mgvi & $  7.1662\times 10^{-5} $\\
Z\tablenotemark{b}      & $  5.9030\times10^{-4}   $\\
\enddata
\tablenotetext{a}{Corresponds to ``standard'' in \tab{tab:abundances_pops}.}
\tablenotetext{b}{Corresponds to [Fe/H] $= -1.7$ for the
  $\alpha$-enhanced mixture considered here.}
\end{deluxetable}

For the mixing-length parameter we adopt $\alpha_\mathrm{MLT}=1.73$,
which is obtained from a standard solar model.  MESA is executed in
hydrodynamic mode and sufficient artificial viscosity is assumed in
order to damp out the large velocities that are otherwise predicted to
occur near the surface during the advanced stages of AGB thermal-pulse
(TP) evolution (especially during the dredge-up phase). In addition to
the default mesh-point controls, we refine the mesh using the
abundances of \p, \hevi, \cdr\ and \nvi\ so that the chemical
abundance profiles are always well resolved.  Additional criteria have
been implemented in order to improve the resolution of He-shell
flashes and the advance of the thin H-burning shell during the
interpulse phase as a function of time.  We use the atmosphere option
\texttt{'simple\_photosphere'}, and we assume a mass-loss rate of
$\log\mdot\ \mathrm{(in\msun/\jahre)}\approx -4.8$ to $-4.3$, which is
obtained using the \citet{bloecker:95a} mass-loss rate that has been
incorporated in MESA with $\eta = 5\times10^{-4}$. (Although some
sequences were followed for up to 100 TPs, none of the model stars had
lost significant amounts of mass by the time the calculations were
terminated.)  We also adopt the OPAL opacities that include C- and
O-enhanced tables \citep{iglesias:96}. In the pre-AGB evolution we
consider convective boundary mixing (CBM) according to the exponential
CBM model \citep{freytag:96,herwig:97,herwig:99a}, with $f=0.014$ at
all convective boundaries.

We have chosen initial masses of $5.0\msun$ and $7.0\msun$ to
represent, in turn, an intermediate-mass AGB star having a CO core and
a hydrogen-free core mass of $\mh=0.991\msun$ at the first TP, and a
super-AGB star with an ONe core and a hydrogen-free core mass of
$\mh=1.2677\msun$ at the first TP.  We present below the results that
have been obtained for these two cases when both standard and modified
assumptions are made about stellar interior mixing processes.

\subsection{Intermediate-mass AGB star model}
\subsubsection{Standard mixing assumption}
\label{sec:standard-5m}
We will first describe the surface abundance evolution of the standard
5\msun\ calculation and how it relates to the evolution of convection
zones in some detail (\abb{fig:surf_abu-5m}). For the standard case, we
assume convective boundary mixing with $f_\mem{ce}=0.004$ at the
bottom of the convective envelope and $f_\mem{pdcz}= 0.002$ at the
bottom of the pulse-driven convection zone.  Both of these values are much
smaller than those normally assumed in computations of low-mass AGB
models. As shown by \citet{herwig:04c} a CBM parameter as large as
indicated by \spr-constraints in low-mass AGB stars
($f_\mem{ce}\sim0.13$) would lead to very vigorous hot dredge-up which
eventually evolves into a corrosive flame penetrating deeper and
deeper into the core. The model evolution is soon terminated in this
case due to high mass loss induced by extreme luminosities. Such a
scenario seems incompatible with the requirements for the ejecta of
intermediate-mass and super AGB stars in \ocen. This issue does,
however, highlight shortcomings regarding our present models of CBM in
the deep stellar interior.
\begin{figure}[tbp]
   \includegraphics[width=0.48\textwidth]{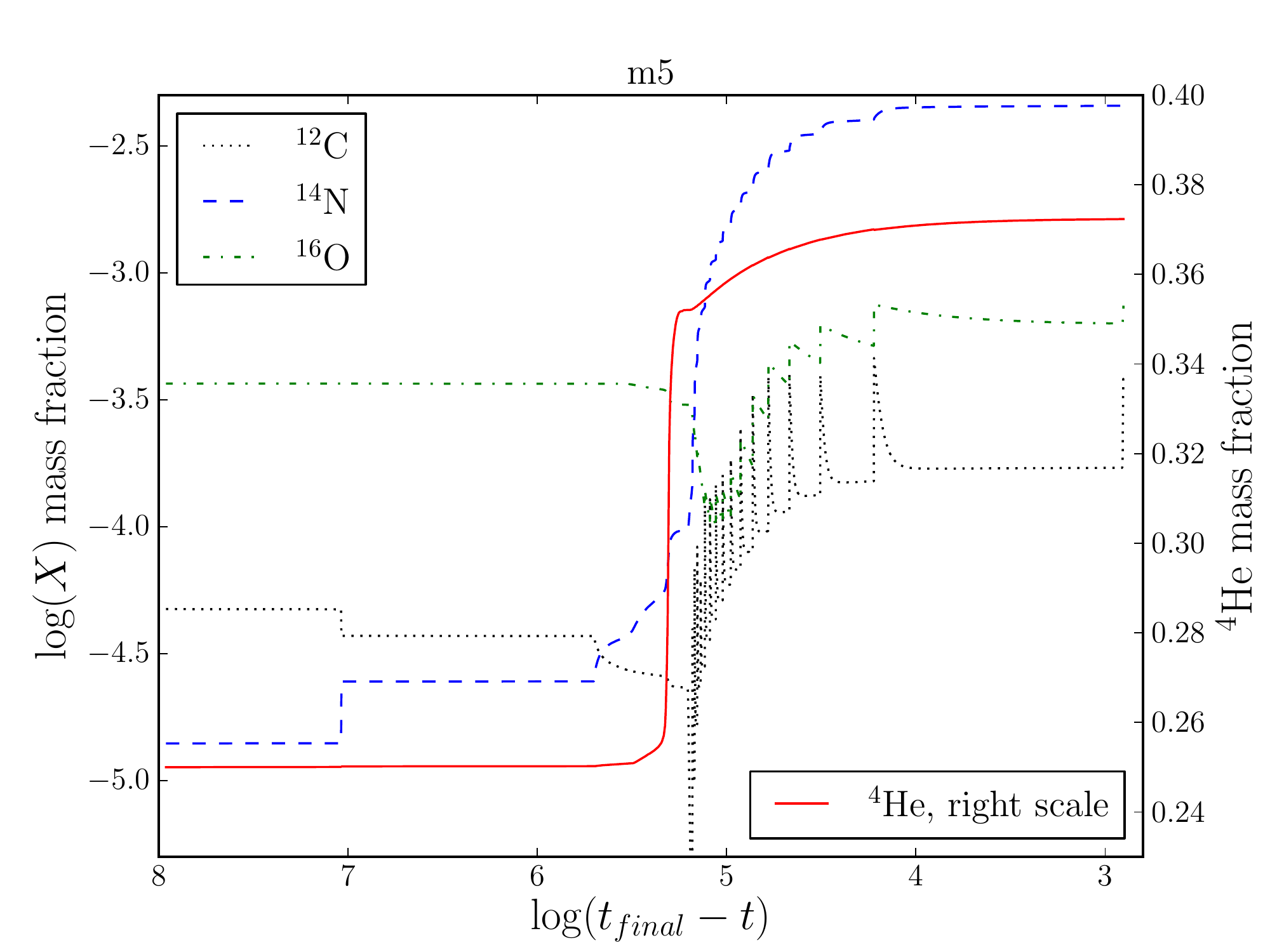}
   \includegraphics[width=0.48\textwidth]{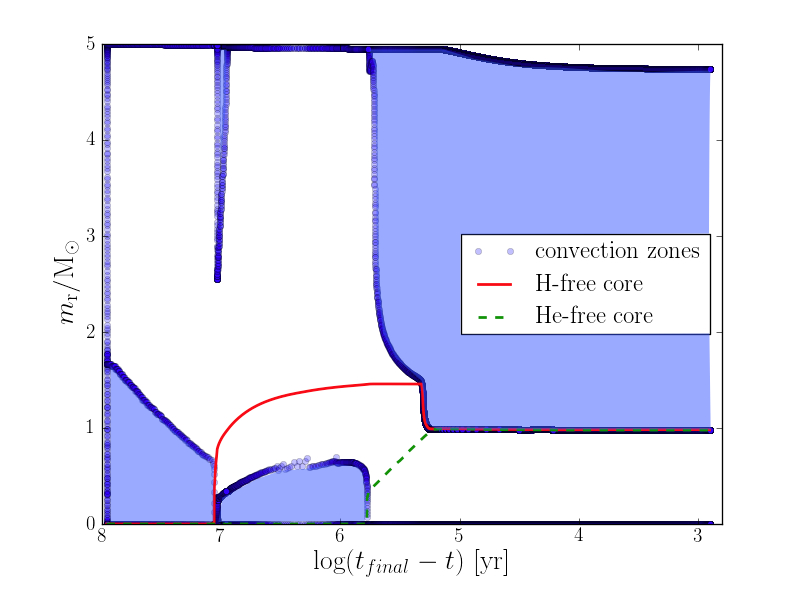}
   \caption{Surface abundance evolution of dominant CNO isotopes and
     \hevi\ of 5\msun\ stellar evolution tracks with standard mixing
     assumptions (top panel) and Kippenhahn diagram (bottom
     panel). The time axis is reverse-logarthmic to accomodate the
     accelerating evolution, $t_\mathrm{final}=\natlog{9.077424}{7}\jahre$.}
\label{fig:surf_abu-5m}
\end{figure}

At the beginning of the evolution all chemical species have surface 
abundances that are unmodified from those listed in \tab{tab:don_A09C},
and they do not change until the end of H-core burning. Then, at
$\log (t_\mem{final} -t) = 7$, the first dredge-up leads to a small
decrease of the surface \czw\ abundance and a corresponding increase
of the surface \nvi\ abundance. No changes are predicted during He-core
burning, after which major changes, again initially for \czw\ and \nvi\ 
but then for \hevi. They are the result of the second dredge-up, which
starts at $\log (t_\mem{final} -t) = 5.7$ and ends at $5.25$. The second
dredge-up, which takes place even before the first thermal pulse on
the AGB, mixes into the envelope a large fraction of the He that was produced
by H-shell burning during the He-core burning phase. This mixing event
is responsible for a major increase in the surface He abundance, from
initially $X(\hevi)=0.250$ to $0.353$ for this case. By the end of the
second dredge-up the \nvi\ abundance has increased by $+0.6 \mem{dex}$,
\czw\ has decreased by $0.3 \mem{dex}$ and \ose\ has decreased by
$0.08 \mem{dex}$.

These modifications of the surface, and therefore envelope
composition, all point in the right direction, but quantitatively they
are far from matching the ``extreme" abundances listed in
\tab{tab:abundances_pops}.  However, further processing of the envelope
takes place during the thermal-pulse AGB phase that starts shortly
after the end of the second dredge-up. Individual TP-related mixing
events, like the third dredge-up and the He-shell flash convection
zones are fully resolved in the stellar evolution calculation, but not
in the Kippenhahn plot shown in \abb{fig:surf_abu-5m}. However, for the
5\msun\ case, the repeated action of the third dredge-up after each
thermal pulse is evident from the periodic step-like increase of the
\czw\ and \ose\ surface abundance evolution, which is also shown in
\abb{fig:surf_abu-5m}. The thermal-pulse AGB evolution starts at $\log
(t_\mem{final} -t) = 5.2$, and though little or no third dredge-up
occurs during the five He-shell flashes, there is a steep initial
increase of the \nvi\ envelope abundance up to a mass fraction of
$\log(X) = -3.0$ at $\log (t_\mem{final} -t) \sim 5.1$.  This is due,
at first, to the destruction of \ose\ when the third dredge-up is weak,
and then to just the effects of deep third dredge-up (typically
$\Delta m_\mem{dup}=1.7\times10^{-3}\msun$ per pulse) after each of the
next six TPs,  These events mix into the envelope primary \czw\ and
\ose\ from He-shell burning.  Note that, due to the hot dredge-up
in our models \citep{herwig:03c}, the immediate CN-cycling conversion
of \czw\ to \nvi\ during the dredge-up phase is responsible for
approximately $20\%$ (a fraction that decreases to below $10\%$ in
subsequent thermal pulses) of the \nvi\ production per pulse cycle
(also see \abb{fig:prof_5m_13tp} and the discussion below).
Hot-bottom burning during the interpulse phase is responsible for the
remainder of the \nvi\ production from the transmutation of both
\czw\ and \ose\ via the CNO cycle at the bottom of the convective envelope.
\begin{figure}[tbp]
   \includegraphics[width=0.48\textwidth]{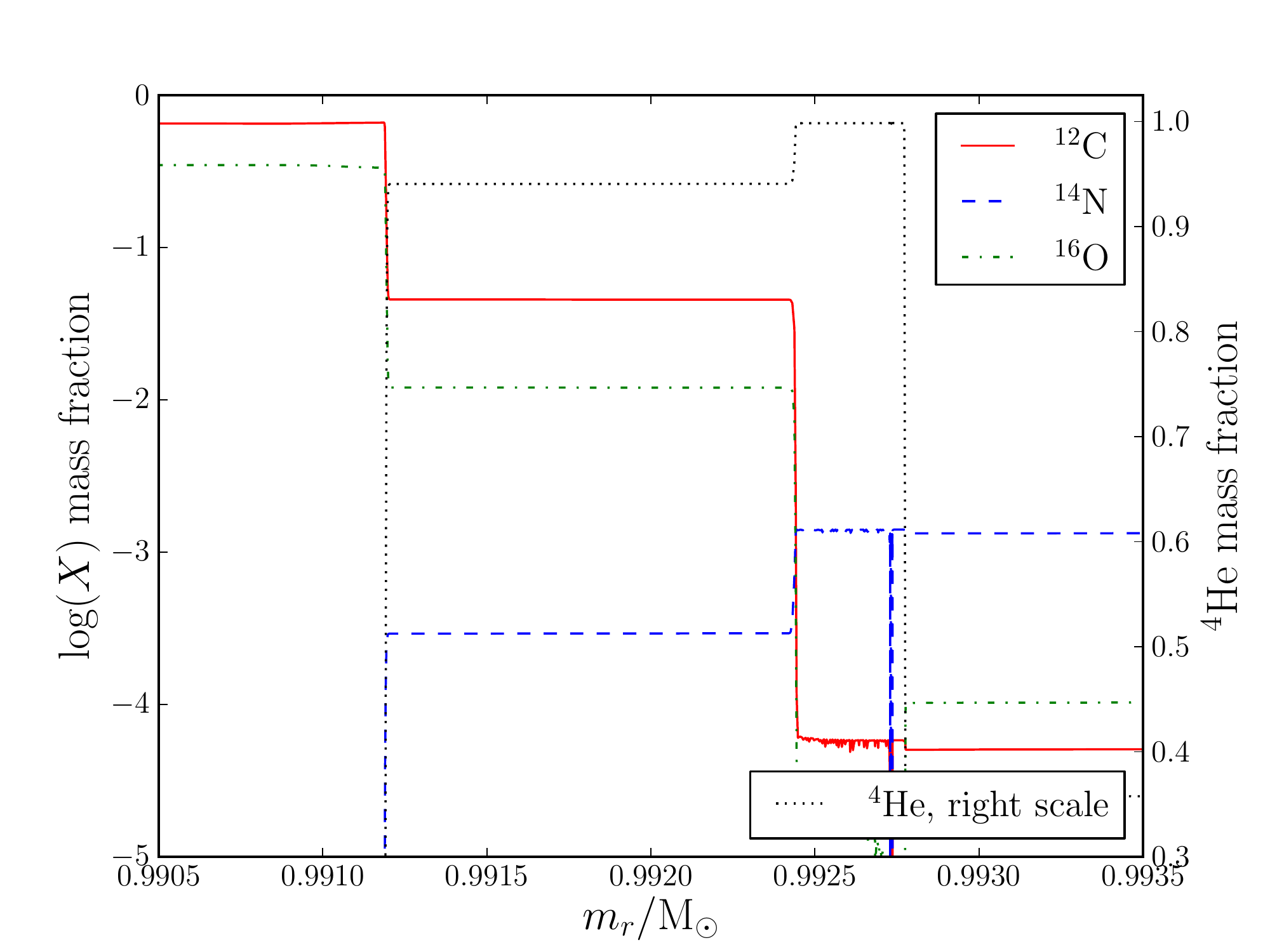}
   \includegraphics[width=0.48\textwidth]{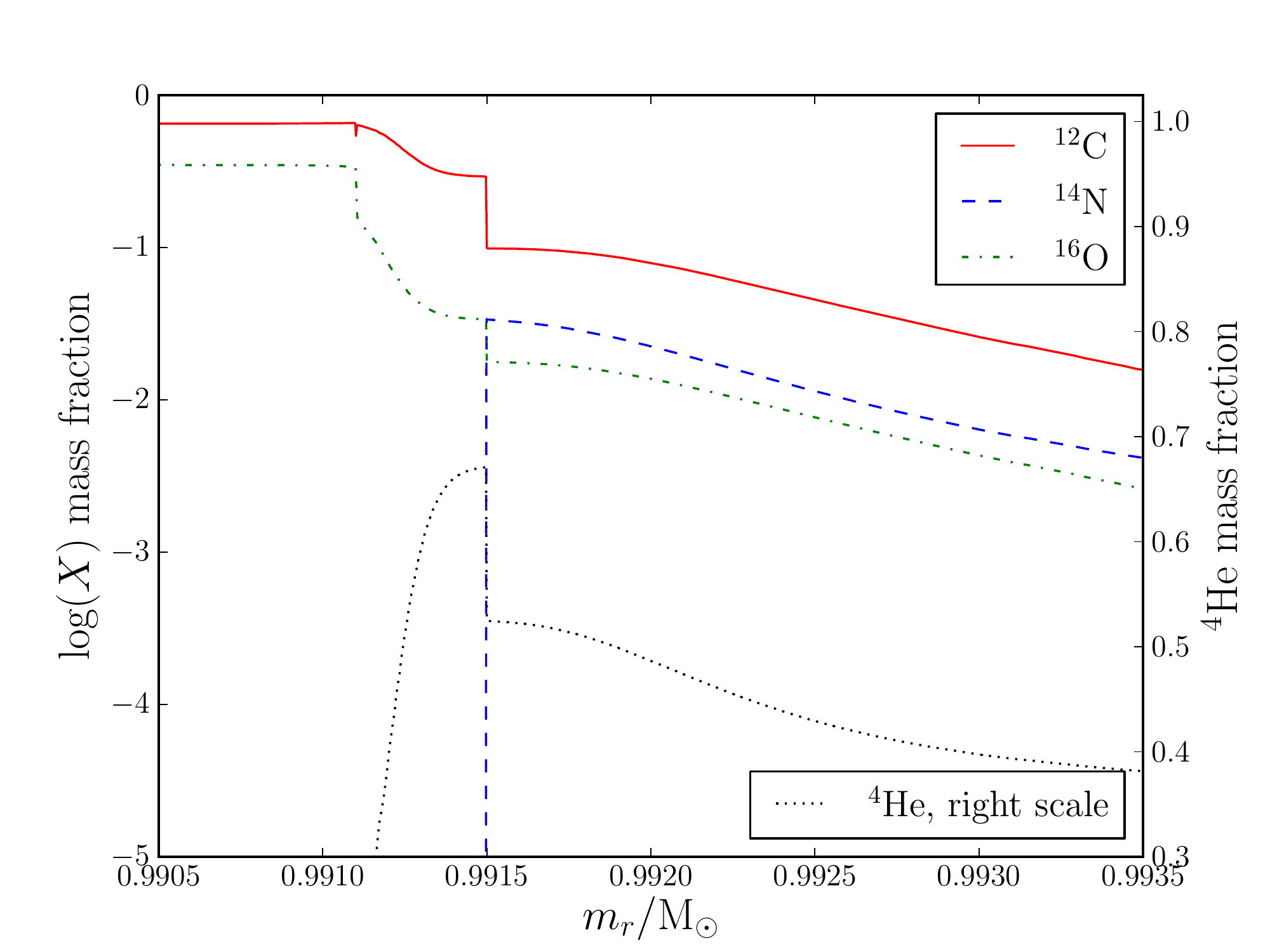}
   \caption{Profile of CNO isotopes and \hevi\ during the He-shell
     (approximately at the end of the first $1/3$ of the duration of
     the He-shell flash convection when the convection zone is still
     growing outward in mass) and during the third dredge-up (when
     $1.3\times10^{-3}\msun$ out of a total of eventually
     $1.9\times10^{-3}\msun$ have been dredged-up) following
     the He-shell flash.}
\label{fig:prof_5m_13tp}
\end{figure}

At $\log (t_\mem{final} -t) = 5.08$ after eleven TPs, of which the last
six were followed by efficient dredge-up, the total \nvi\
enhancement has risen to $1.83\,\mem{dex}$ and the envelope He abundance
is $Y=0.356$. While the \nvi\ abundance is now in reasonably good agreement
with the estimate in the ``extreme" abundance mixture, the value of $Y$ is 
just reaching the lower limit of the range that has been inferred from 
the isochrones. Furthermore, the O abundance has decreased by only $\sim
0.5\,\mem{dex}$ and the C abundance has increased in total by $\leq
0.3\,\mem{dex}$.  These reductions are both significantly less than the
values that define the ``extreme" abundance distribution.  However, for
carbon, in particular, it should be kept in mind that the ``extreme"
composition is based in part on the measured
%\texttt{[need to discuss with Don]}
abundances in RGB stars, which are known to experience extra-mixing
processes \citep[e.g.][]{denissenkov:03b} that cause the C
abundance to decrease with increasing luminosity as they climb the
giant branch \citep{gratton:00}. 

In the more advanced pulses (starting around the tenth TP), each
dredge-up event will mix the entire intershell helium as well as a small
layer interior to the bottom of the He-shell flash convection zone
where $X(\czw)=0.63$ and $X(\ose)=0.35$.  The dredge-up into and below
the former He-shell flash convection zone is shown in
\abb{fig:prof_5m_13tp}. The He-shell flash the convection zone
combines \czw- and \ose-rich material from deeper layers
($<0.9912\msun$) with the \nvi-rich H-shell ashes (between $0.9924$
and $0.9927\msun$). At the time that the profiles are shown the
convection zone has not yet reached its full extent, which occurs when the
upper boundary has approached the location of the former H-burning
shell ($0.9927\msun$). At this point the \czw\ and \ose\ mass-fraction
abundances in the He-shell flash convection zone are $X(\czw)=0.046$ and
$X(\ose)=0.012$, which is considerably lower than the typical
intershell abundances in lower mass models.

After the end of the He-shell flash, the dredge-up (see the bottom
panel of \abb{fig:prof_5m_13tp}) proceeds into the H-free core.
The profiles illustrate how \nvi\ is produced {\it in-situ} as
\czw\ streams through the hot convective boundary at
$m_\mem{r}=0.9915\,\msun$. In the outward direction, it is the
simultaneous burning of \nvi\ and the convective mixing in the envelope
that causes the sloped chemical abundance profiles. The \hevi\ abundance
just below the convective boundary is reduced from the value obtained
at earlier times (see the top panel) due to nuclear burning, which
forms a small additional amount of \czw\ and \ose\ to be
subsequently dredged-up. However, close inspection of the
models shows that the bulk of the \czw\ and \ose\ that is mixed to the
surface does not come from the region occupied by the former
convective He-shell flash, but originates instead from the layer just
beneath. The deepest mass coordinate reached by the third dredge-up in
this event is $0.9908\,\msun$. Therefore, the amount of \czw\ coming from
the region below the former He-shell flash convection zone is
$0.0003\msun\ \times 0.63 = 1.8 \times 10^{-4}\msun$, which is
approximately the same as the \czw\ content of the envelope at this
point. Thus, the visible spikes in the \czw\ surface abundance
evolution (\abb{fig:surf_abu-5m}) and much of the overall enhancement
of CNO is due to deep dredge-up that reaches below the He-shell
flash convection zone.

This deep dredge-up will also bring \ose\ from below the He-shell
flash convection zone into the envelope. In this model the dredge-up
of \ose\ outperforms the capacity of hot-bottom burning to reduce the
\ose\ abundance in later TPs. As a result, this model predicts an
increasing \ose\ abundance, which would eventually lead to an enhanced
O abundance in the AGB wind ejecta compared with the initial O abundance.
This is clearly in contradiction to the target abundance in the
``extreme" mixture.  While this deep dredge-up will also mix all of
the He in the helium shell into the envelope, this contribution to
the evolution of the envelope helium abundance is
negligible.  The bulk of the He comes from hot-bottom burning during
the interpulse phase. During the 15 fully developed thermal pulses
(i.e., not counting the first five TPs that have insignificant
dredge-up) the He mass-fraction abundance increases by
$\Delta Y = 0.02$, reaching $0.372$ in the last computed model.
Therefore, a value of $Y$ of $\approx 0.4$ could be obtained after
another 21 thermal pulses, which is in all likelihood a very
plausible scenario. 

In conclusion, our standard intermediate-mass AGB stellar model is
able to generate wind ejecta with He as well as N abundances in the
ranges required by the ``extreme" mixture. As a matter of fact, the
total N enhancement according to our model is $2.5\,\mem{dex}$, which
exceeds the enhancement specified in the ``extreme'' abundance
distribution.  However, it is possible to alter the abundances of C, N
and O in the models by fine tuning (reducing) the convective boundary
mixing (CBM) parameter at the bottom of the convective envelope. This
may be necessary, in fact, given that there will be many more thermal
pulses in a real star than we have computed here, with the consequence
that the envelope enrichment would be stretched out over a much larger
number of dredge-up events.  However, because the \czw\ for the
production of \nvi\ during hot-bottom burning and \ose\ come from the
same region in the star, it is not possible to obtain a significant N
enhancement and, at the same time, a significant O reduction (to be
consistent with the ``extreme" abundance distribution) without further
modifications of our model assumptions.

\subsubsection{Increased mixing-length parameter}
\label{sec:increased_amlt}
In addition to boundary mixing uncertainties, some assumptions need to
be made about the mixing-length parameter, \amlt, when modeling
convection in the envelopes of AGB stars.  It is commonly assumed that
the value of \amlt, as calibrated by fits to the solar parameters, can
be applied universally to all convection zones during all evolutionary
phases. However, both simulations as well as semi-empirical evidence
suggest that the deep convection zones of giant envelopes, comprising
essentially fully convective configurations, are better described by
an increased mixing-length parameter.  A variable (i.e., non-constant)
\amlt\ parameter was already suggested by the radiation-hydrodynamics
simulations by \citet{ludwig:99}, also see the later work by
\citet{robinson:04}.  Specifically relevant to giant stars,
\citet{porter:00a} presented 3D-simulations of deep envelope
convection that were best reproduced within the mixing-length picture
if $\amlt=2.68$, assuming the formulation of the MLT given in
\citet{cox:68}.  In addition, a semi-empirical determination of \amlt\
was obtained from the the modeling of the pulsation of highly-evolved,
variable, intermediate-mass stars ($4-6\msun$) in the LMC and SMC by
\citet{mcsaveney:07} when they derived spectroscopic abundances.  For
three giants, they determined significantly larger values of \amlt,
ranging from 2.2 to 2.4 --- also based on the \citet{cox:68} version
of the MLT. Taken together, there is ample justification to explore
the effects of assuming a larger value of \amlt.  Numerical
experiments with enhanced values of \amlt\ have been carried out for
intermediate mass AGB stars and super-AGB stars previously by
\citet{Karakas:2012kc} and \citet[][, using a post-processing
code]{siess:10}.

To investigate this issue, we have computed two evolutionary sequences
starting at the first thermal pulse of the standard sequence that was
described in the previous section. For the convective boundary mixing
(CBM) parameters, we used $f_\mem{ce}=0.002$ and $f_\mem{pdcz}=0.001$,
while at all other convective boundaries, we adopted $f=0.01$.  One
sequence had $\amlt=1.73$, as assumed in the standard sequence, and
the other had $\amlt=2.40$.  Both sequences were followed through the
initial 5 to 6 TPs with no or little third dredge-up, and the
computations were halted after another two full TP cycles with fully
developed third dredge-ups had been completed. The evolutionary
properties of the models over those two TP cycles are compared in
\tab{tab:amlt}. The $\amlt=2.40$ calculations show, on average, higher
interpulse H-burning, and peak-flash He-burning, luminosities. For the
high-\amlt\ case, the third dredge-up is more than twice as deep, and
it reaches below the bottom of the convection zone boundary
established by the previous He-shell flash at the second fully
developed dredge-up, as compared with the standard-\amlt\ case wherein
the dredge-up proceeds into about two-thirds of the former He-shell
convection zone. This standard-\amlt\ case predicts a shallower third
dredge-up compared with the sequence described in
\kap{sec:standard-5m} because the adopted value of $f_\mem{ce}$ is
smaller.

The convection assumptions, primarily parameterized through \amlt\ and
$f_\mem{ce}$, significantly alter the evolution of the envelope
abundances.  Over the first two pulses with significant dredge-up, the
enhancement in the He abundance is a factor of $6.6$ higher for the
$\amlt=2.40$ case than in the models for $\amlt=1.73$.  Also, in the
$\amlt=1.73$ case, the increase in $\Delta X(\hevi)$ per pulse at
later times is consistent with the average value for $\Delta X(\hevi)$
reported for the standard case in the previous section.  Since, in a
differential sense, the same is true for the $\amlt=2.40$ case, we
find that intermediate-mass AGB models with a larger value of the
mixing-length produce He from hot-bottom burning more efficiently.
Insofar as the CNO elements are concerned, the deeper dredge-up that
occurs when $\amlt=2.40$ means that significantly more C and O is
brought into the envelope, even from below the former He-shell flash
convection zone. However, the more efficient hot-bottom burning in the
model with higher \amlt\ means that, not only is N produced
efficiently, but even O is depleted efficiently. Over the first two
pulses of the $\amlt=2.40$ sequence, the O abundance is depleted by
$\sim -0.6\dex$ in comparison with the initial abundance.

\subsubsection{Conclusion}
\label{sec:agb-concl}
We conclude that, with the right combination of the convection model
parameters \amlt\ and $f_\mem{ce}$, it should be possible, in
principle, to account for He mass-fraction abundances up to $Y\sim
0.40$, as well as sufficiently enhanced N and depleted O, in the wind
ejecta from intermediate-mass AGB stars.  Indeed, to obtain the low
oxygen abundances that have been observed, it seems to be necessary to
adopt a high value of \amlt.  On the other hand, the increased
effectiveness of third dredge-up that is also found when a larger
value of \amlt\ is assumed, may lead to an enhancement of N that is
too large for a given increase in $Y$.  We have therefore constructed
one additional $5\msun$ stellar model sequence with $\amlt=2.4$, but
with a further reduction in the value of the $f_\mem{ce}$ parameter.
This case has been evolved over ten thermal pulse cycles with fully
developed dredge-up (see the top panel of
\abb{fig:surf_abu-7m}). While the N enhancement now reaches
$\sim2\dex$ and the He abundance reaches the average value of the
range that is indicated by fitting isochrones to the bMS, the O
abundance has been depleted by about $1\dex$, just as required by the
``extreme" abundance distribution. However, even this model sequence
does not show a significant C depletion.
\begin{figure}[tbp]
   \includegraphics[width=0.48\textwidth]{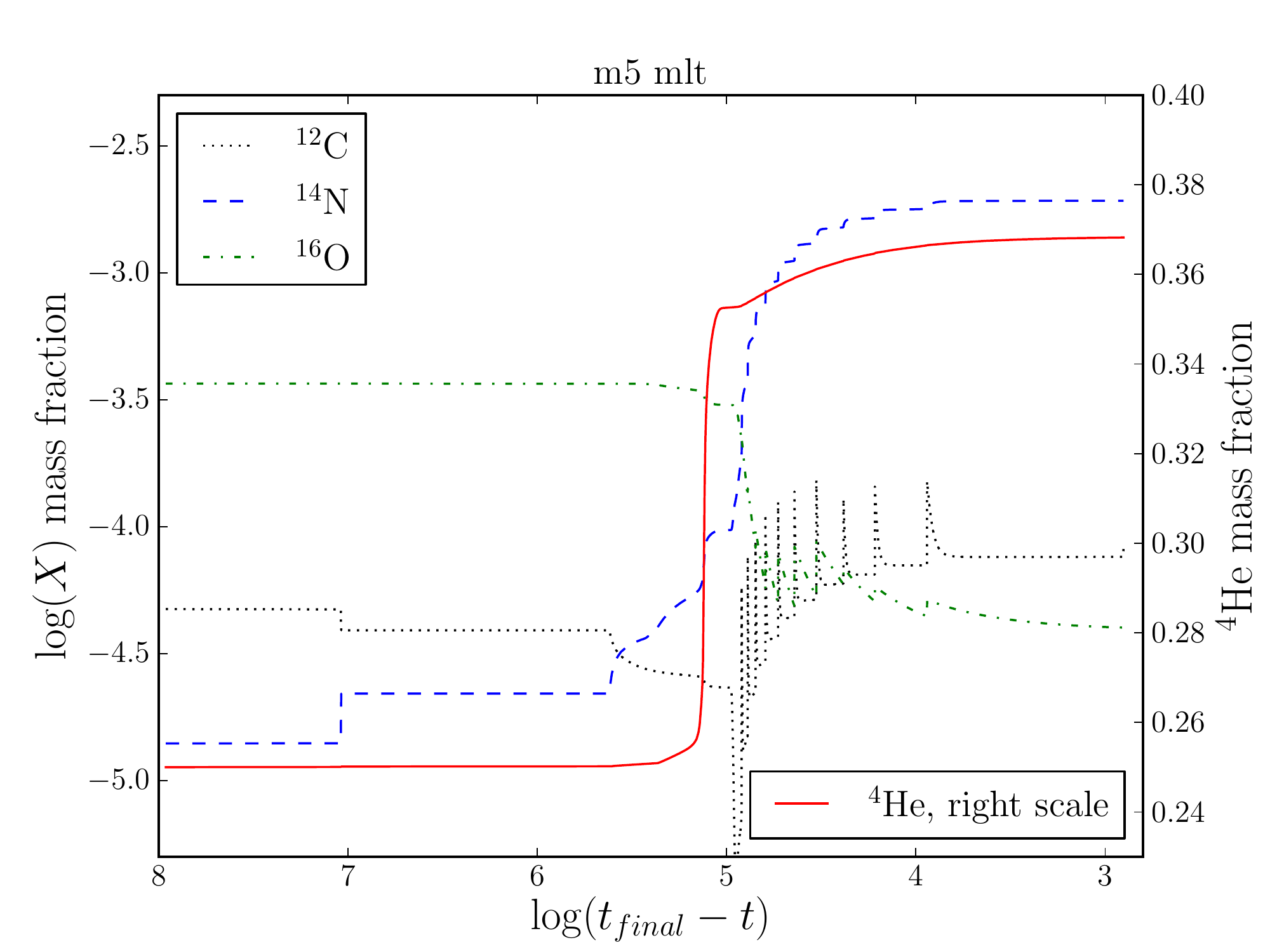}
   \includegraphics[width=0.48\textwidth]{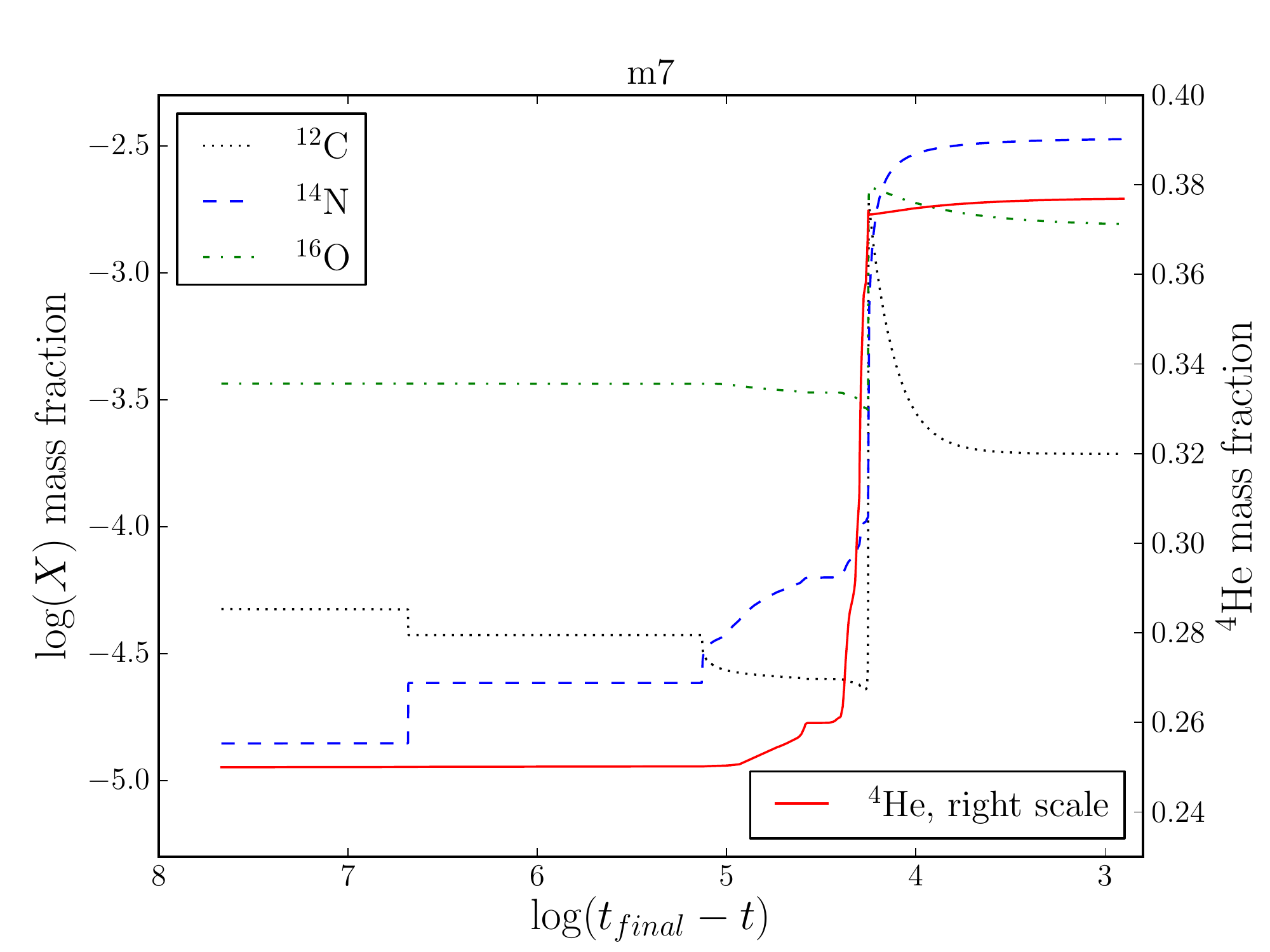}
   \caption{Surface abundance evolution of 5\msun\ stellar evolution
     track with enhanced mixing-length parameter $\amlt=2.4$ (top
     panel, $t_\mathrm{final}=\natlog{9.1641578}{7}\jahre$) and
     surface abundance evolution for 7\msun\ track with standard
     mixing assumption (bottom panel,
     $t_\mathrm{final}=\natlog{4.6642752}{7}\jahre$).}
\label{fig:surf_abu-7m}
\end{figure}

A real intermediate-mass AGB star will likely do more thermal pulses than we
were able to compute here. This means either that the real star may
have convection properties that will stretch out the chemical abundance
modifications seen in our stellar models over more thermal pulses, or that
the ejecta of the AGB star are more extreme in their abundance patterns
than indicated by the ``extreme'' metals mixture, which would leave
some room for the dilution of the ejected winds with unprocessed ``standard"
abundances before forming the bMS generation of stars. As discussed
earlier the ``extreme'' abundance mix may in fact be an upper limit
for the bMS in \ocen, and a somewhat less extreme abundance mix in
this cluster would add to the possibility of dilution. 

%% FALK: I have taken this out again, because we now do not believe in
%% dilution anymore. I need to update the response to the referee.
%It is well known that AGB star models can not generate the observed
%Na-O anticorrelation
%\citep[e.g.][]{denissenkov:03a,D'Antona2011}. Dilution with primordial
%gas flowing into the cluster may solve this problem
%\citep{d'ercole:08}, and it may also diminish the C overproduction
%observed in our models. 

With these caveats in mind, it seems very
reasonable to conclude that intermediate-mass AGB star models are
capable of producing abundance patterns very similar to that listed in
\tab{tab:abundances_pops} under the column heading ``extreme".

\subsection{Super-AGB model}
\label{sec:sagb-models}
For somewhat higher initial masses, stars have a core mass at the
first thermal pulse which is larger than approximately $1.05\msun$, in
which case, C burning is ignited and those stars will form a ONeMg
core \citep{garcia-berro:94}. These are the so-called ``super-AGB" stars, which, like their
intermediate-mass AGB star counterparts with CO cores, have periodic
thermal pulses, dredge-up events, and hot-bottom burning. It has been
suggested that these super-AGB stars could provide the chemically
peculiar ejecta that are required in order for a stellar population
like the bMS in \ocen\ to form.

\subsubsection{Deep second dredge-up mixing in the  $7\msun$ case}
\label{sec:deep2dup}
We have calculated stellar
evolution sequences with an initial mass of $7\,\msun$ that adopt the same
initial abundances, and employ the same physical assumptions, as in
the $5\msun$ sequence discussed above, but explore uncertainties in
convection assumptions.

In \abb{fig:surf_abu-7m} we show the evolution of the surface
abundances for our standard case.  As reported for the lower mass
models in \kap{sec:standard-5m}, the chemical composition changes as a
result of the first dredge-up (at $\log (t_\mem{final} -t) = 6.7$),
while the second dredge-up begins to have an impact at $\log
(t_\mem{final} -t) = 5.1$.  The He abundance gradually increases as
the bottom of the convective envelope penetrates into the ashes of the
H-burning shell.  In some super-AGB models, depending on mass (see
\kap{sec:inimass}) and details of the physics assumptions, a He-shell
flash convection zone has developed before the end of the second
dredge-up. The quenching of the flash causes a further dredge-up, also
refered to as a ``dredge-out" event \citep{iben:97,ritossa:99,siess:07}. This
final mixing episode is shown in more detail in
\abb{fig:dredge-out}. Depending on the mass the second dredge-up
proceeds into the He-free core. This mixing into the He-free core adds
additional C, and later on also O to the envelope, as seen in
\abb{fig:dredge-out} and \abb{fig:surf_abu-7m} at $\log (t_\mem{final}
-t) = 4.25$.  This dredge-out mixing increases the envelope O
abundance in our $7\,\msun$ model by $\sim 0.77\dex$, which is
completely at odds with the $\sim 1\,\dex$ reduction that is required
in order to match the abundance given in the target ``extreme"
mixture.  The C abundance is also significantly enhanced, by
$1.6\dex$, with respect to the intial value.
\begin{figure}[tbp]
   \includegraphics[width=0.48\textwidth]{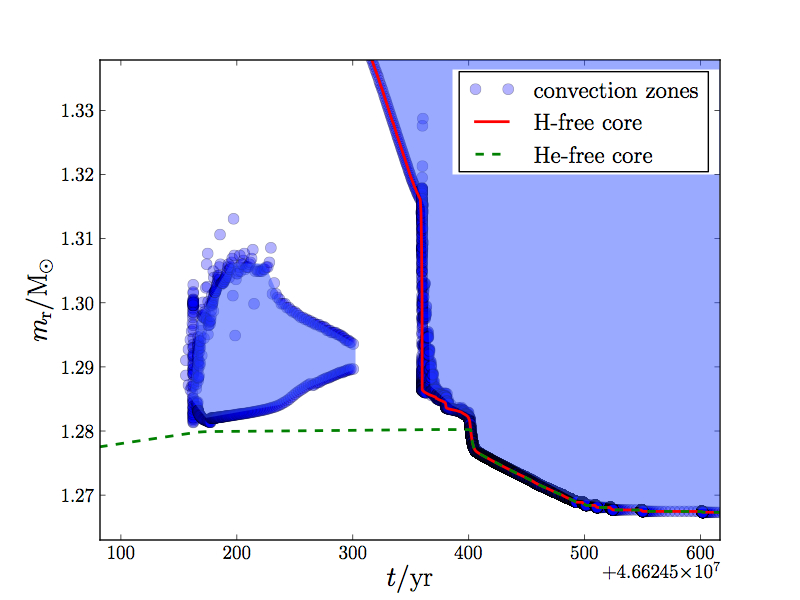}
   \includegraphics[width=0.48\textwidth]{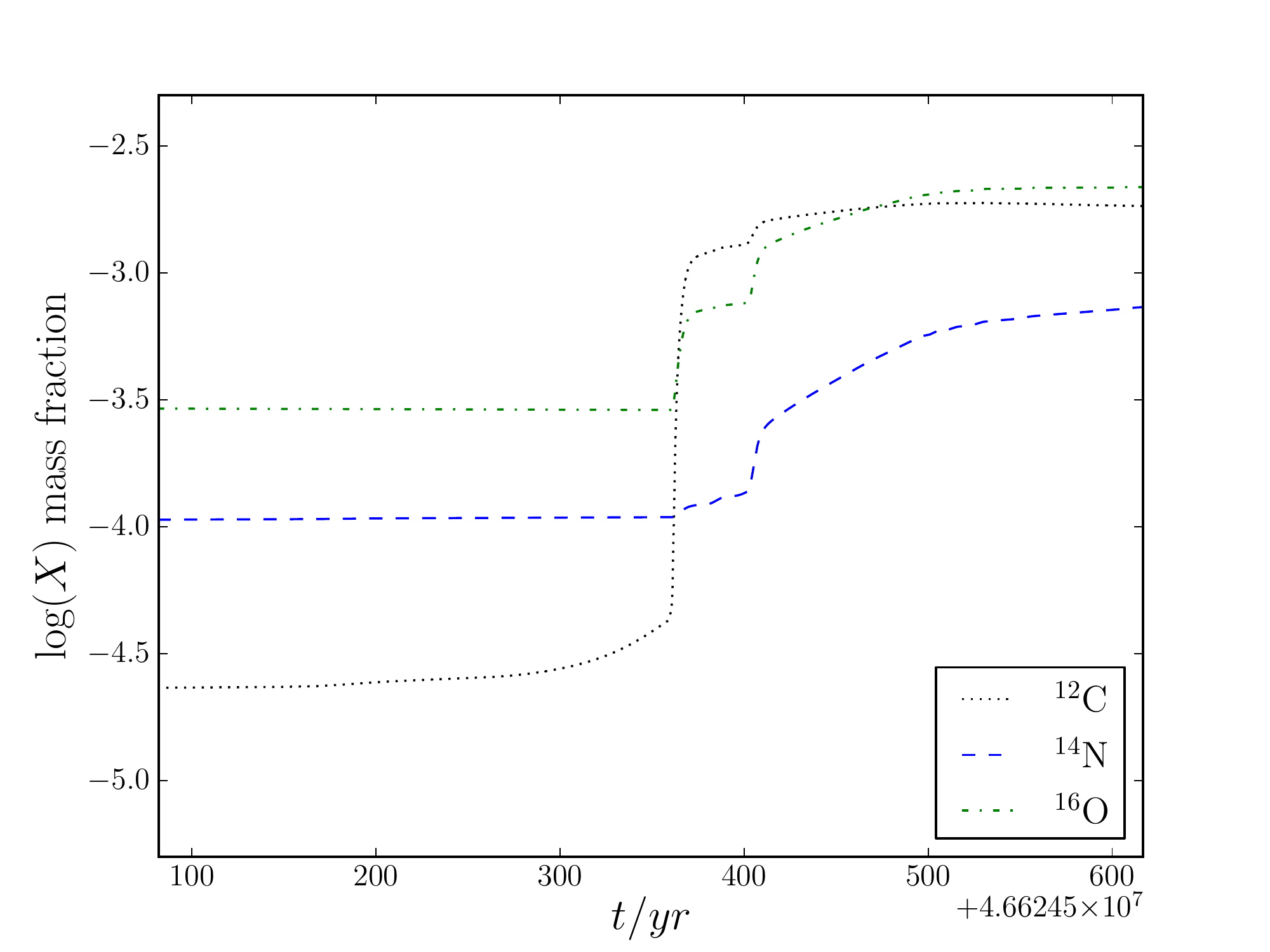}
   \caption{Top panel: detail of the last phase of the second
     dredge-up of the \mzams=7\msun\ case; bottom panel: simultaneous
     surdace abundance evolution. After a so-called
     ``dredge-out'' phase the bottom of the core convection zone
     continues to penetrate (on Lagrangian coordinate) into the
     He-free core, bringing first \czw\ and then \ose\ into the
     envelope. The quantitative details of these mixing processes are
     sensitve to uncertain physics assumptions (see text). H- and
     He-free boundaries correspond to the mass coordinate where the
     mass fraction is respectively $\leq 10^{-4}$.}
\label{fig:dredge-out}
\end{figure}

One may think that efficient hot-bottom burning in super-AGB stars,
possibly boosted by the adoption of a higher value of \amlt\ could reduce
even this additionally dredged-out O and C.  Although individual third
dredge-up events cannot be identified in the plot which shows the evolution
of the envelope abundances, the $7\msun$ sequence has been evolved
through 18 thermal pulses with deep dredge-up.  They cause the helium
abundance to increase by $\Delta X(\hevi) =0.0002$ per TP, which is
the same as that found in the standard MLT sequence presented in
\tab{tab:amlt}. In order to reach a He mass-fraction abundance of $0.40$
another $105$ thermal pulse cycles would be needed, which is a very
likely possibility considering the short interpulse periods and a reasonable
range of possible mass-loss rates. 

During the same initial 18 thermal pulses, the O abundance decreases by
$0.16\dex$; consequently, over another $\sim 100$ TPs, a significant O
depletion could be achieved. However, even if hot-bottom burning is able
to further increase the abundance of He, and reduce the O abundance, by
the desired amounts, the N abundance will become far too large.  At the
end of 18 thermal pulses, the \nvi\ abundance has already increased
by more than $2.3\dex$ and futher depletions of O will only add to the
production of N. While the total CO enrichment in CO-core AGB stars can be
controlled by fine-tuning the convection parameters, this is not
obviously possible given the substantial enrichment of some super-AGB
envelopes during the deep second dredge-up that reaches into the
He-free core.

\subsubsection{Initial mass dependence}
\label{sec:inimass}
The deep second dredge-up depends on assumptions of convective
boundary mixing a well as on stellar mass.  In order to determine the
mass dependence of this mixing, as well as the upper and lower
limiting masses for the super-AGB phases and the lifetimes for
intermediate-mass AGB and super-AGB stars, we have calculated a fine
grid of models with $4.0 \leq \mzams/\msun \leq 8.80$ with $\Delta
\mzams = 0.2\msun$. Initial abundances are again according to the
standard mix, and convective boundary mixing for H- and He-core
burning with $f=0.014$ was included. At all other convective
boundaries $f=0.010$ was assumed. For this grid v.\,3372 of the MESA
stellar evolution code was used, together with the NACRE
\citep{angulo:99} reaction rates.  From these calculations we find the
lifetimes shown in \abb{fig:mini-starage}.
\begin{figure}[tbp]
   \includegraphics[width=0.48\textwidth]{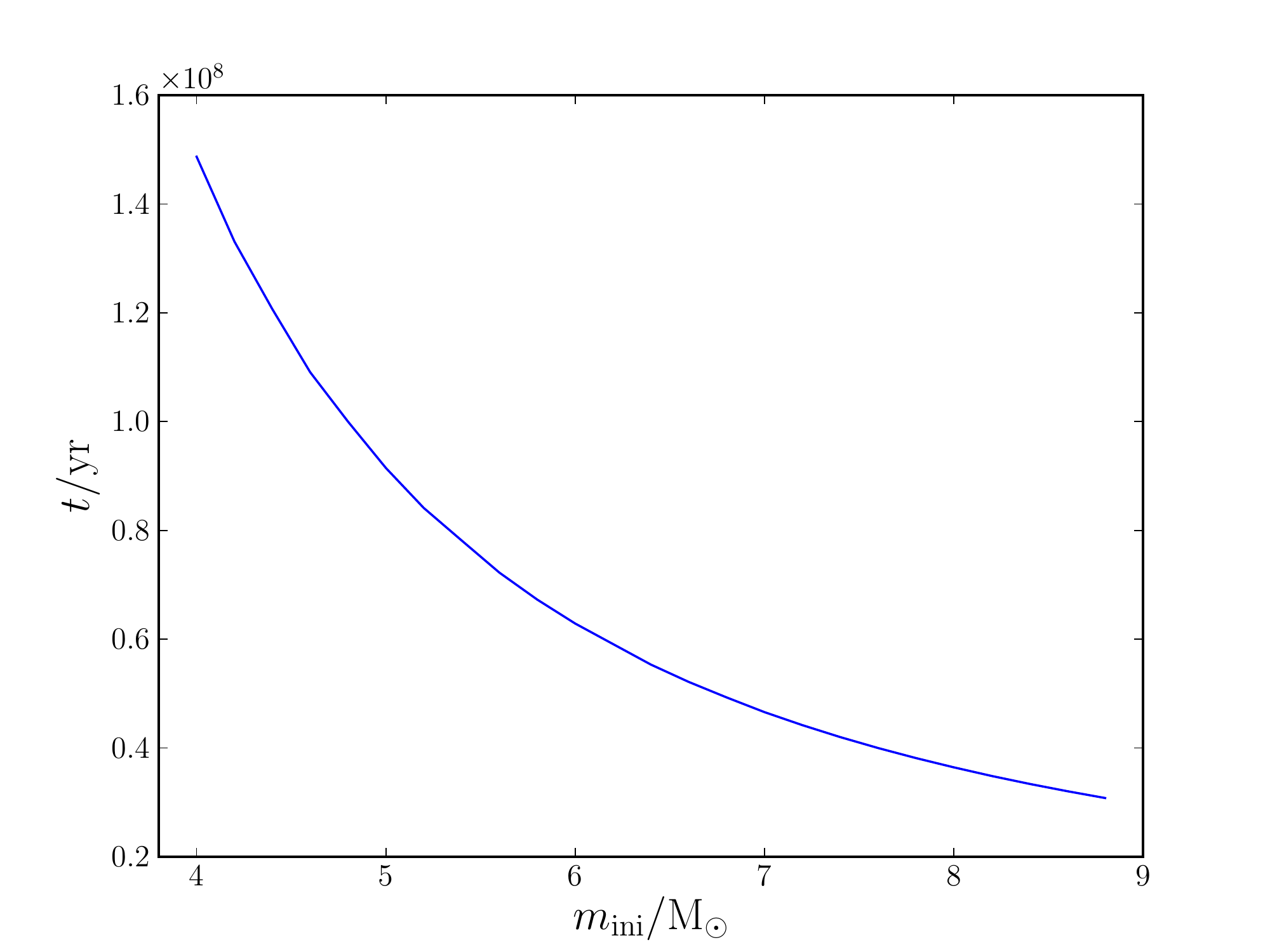}
   \caption{Lifetimes of intermediate-mass AGB and super-AGB for the
     ``standard'' initial composition mix given in
     \tab{tab:abundances_pops}, determined from a grid of stellar
     models described in \kap{sec:inimass}.  }
\label{fig:mini-starage}
\end{figure}

For $\mzams \geq
5.6\msun$ with H-free core masses of $\mh\geq1.05\msun$ the models
show off-center C ignition, which in these models does not burn to
completion in the central region in most cases. Deep second dredge-up
that proceeds into the He-free core is found for $\mzams \geq
6.8\msun$, and in these models the maximum He-free core mass before
the reduction through the second dredge-up is $\mhe\geq
1.245\msun$. Above this mass models predict large dredge-up of primary
CNO as described in \kap{sec:deep2dup}. For $\mzams \geq 7.6 -
7.8\msun$ the core mass after the second dredge-up is at or just below
the Chandraskhar mass, and more massive stars are expected to explode
as supernova. 

%The exact nature of the convective
%mixing episodes in stars of transition mass between
%super-AGB and massive stars models that will ignite Ne, O etc.\ and
%eventually collapse as Fe-core collapse supernova, is not well
%understood, and will quantitatively depend on the exact treatment of
%convection, including convective boundary mixing assumptions not only
%for H- and He-core convection but also for post-He core burning
%phases as well as the correct and complete treatment of nuclear
%networks (Jones etal., 2012, in prep).
 
\subsubsection{Convective boundary mixing effects}
We have calculated another $7\msun$ sequence without convective
boundary mixing at the bottom of the convective envelope, but its
behaviour with regard to the yields of He and CNO over the first dozen
thermal pulses is very similar to the standard case. It shows the same
deep second dredge-up. For these more massive super-AGB stars (cf.\
\kap{sec:inimass}) the yields are dominated by the second
dredge-up. The third dredge-up will only double the C+N+O abundance
over the 18 computed thermal pulses (in the form of N), while the
sequence without convective boundary mixing will not add additional
CNO because there is no third dredge-up. 

\begin{deluxetable}{rll}
\tablecolumns{3}
\tablewidth{\columnwidth}
\tablecaption{ Comparison of two thermal pulse cycles with standard
  ($\amlt=1.73$) and increased mixing-length parameter
  ($\amlt=2.40$): difference in H-shell interpulse and He-flash peak
  luminosities, dredge-up depth in \msun\ for second TP with deep
  dredge-up discussed in the text, surface He mass fraction increase over the
  two TP cycles as well as change of surface  C, N and O over the two
  considered TP cycles. 
\label{tab:amlt}}
\tablehead{
\colhead{}&\colhead{\amlt=1.73}&\colhead{\amlt=2.40}}
\startdata
$\Delta \log \lumh [2.40 - 1.73]$ & \multicolumn{2}{c}{$\phantom{-} 0.2$} \\
$\Delta \log \lumhe [2.40 - 1.73]$  & \multicolumn{2}{c}{$\phantom{-} 0.3 - 0.5$} \\
$\Delta m_\mem{dup}/\msun$ &$\phantom{-}\natlog{7.1}{-4}$&$\phantom{-}\natlog{1.66}{-3}$\\
$\Delta X(\hevi)$&$\phantom{-} 0.0004$&$\phantom{-} 0.0023$\\
$\Delta \log X(C)$&$\phantom{-} 0.28 $&$\phantom{-} 0.50 $\\
$\Delta \log X(N)$&$\phantom{-} 0.26 $&$\phantom{-} 0.50 $\\
$\Delta \log X(O)$&$-0.10$&$-0.51 $\\
\enddata
\end{deluxetable}

Indeed, significant differences exist when convective boundary mixing
is taken into account or when it is neglected altogether, in that the
former is accompanied by effective third dredge-up with
$\lambda\apleq1$, while in the latter instance, the third dredge-up is
absent. This has the important consequences that the
cores of models with CBM do not grow, and they are therefore incapable
of reaching the Chandrasekhar mass.  As a result, they will become
white dwarfs instead of evolving into ONeMg core-collapse supernova
--- unless they reside in a binary system, in which case they may
still explode through an accretion-induced collapse \citep{fryer:99}.

Another, potentially even more relevant consequence is the presence of
\ipr-conditions in super-AGB models with CBM
\citep{herwig:12b}. Briefly, we find that dredge-up may -- due to the
short thermal time-scale at the core-envelope boundary -- enter into
the intershell immediately after the peak He-burning luminosity has
been reached, and the He-intershell region is still convectively
unstable. Then, depending on the convective mixing assumptions, protons may
be convectively mixed into He-shell flash conditions, a situation in
which high neutron densities of $N_\mem{n} \sim 10^{15}\mem{cm^{-3}}$
would be present \citep{cowan:77,campbell:10,herwig:10a}. The details
of the \iprn\ in super-AGB stars are still very uncertain, mostly due
to our inability do correctly model the interaction between convective
mixing and rapid, combustion-like burning of \czw\ + protons with the
spherically symmetric assumption of stellar evolution, and 3D
simulations, such as those in \citet{herwig:10a} will be
required. However, within the present uncertainty \citet{herwig:12b}
do consider a significant production of many n-capture elements in
super-AGB stars from \iprn, including La, a possibility. The
implications of this possibilty will be discussed again later
(\kap{sec:final-ONa}).

\subsubsection{Conclusion}
Super-AGB stars with lower initial masses than $\sim 6.8 \msun$ and
consequently lower He-free core masses ($< 1.245\msun$), would not
suffer from the additional C and O envelope enrichment from deep
second dredge-up. Stars in this mass range would in principle behave
similar to the massive (CO core) AGB stars described in
\kap{sec:standard-5m}, and could have ejecta with the abundance
pattern of the extreme distribution. The contribution of the most
massive super-AGB stars must be minor due to the unavoidable and
excessive O and N abundances that would characterize their ejecta. In
order to identify the mass ranges and exact outcomes quantitatively in
a more reliable way, the properties of convection in the deep stellar
interior must be investigated in more detail. 

In addition to the uncertainties in the treatment of convection, mass
loss assumptions will significantly effect final and quantitative
abundance predictions for ejecta \citep{ventura:11}. In our models we
have kept mass loss low in order to study the impact of convection
uncertainties in isolation.

%Hi Falk  ... the following table relates [Fe/H] and Z for the normal
%alpha-enhanced mixture (A09C) that I've been considering in my models.
%            [Fe/H]          Z
%       -1.40        1.177 x 10^-3
%       -1.50        9.351 x 10^-4
%       -1.60        7.430 x 10^-4
%       -1.70        5.903 x 10^-4
%       -1.80        4.690 x 10^-4
%       -1.90        3.726 x 10^-4
%       -2.00        2.960 x 10^-4
%Don

\section{Summary and Discussion}
\label{sect:sum}
In this section we provide a brief summary of our results, and then
suggest a scenario for the formation of discrete populations of stars
in \ocen, with an emphasis on the bMS.

\subsection{The main results of this investigation}
\label{sec:disc-main-results}
Consistent with previous findings, we have found that the location of
the bMS in the CMD of \ocen\ can be reproduced by stellar models only
if they assume significantly enhanced He abundances, in the range
$Y=0.35 - 0.40$.  Spectroscopic observations for giants in this system,
and for one blue main-sequence star in NGC$\,$2808, suggest that it is
not only helium but also the CNO elements that have peculiar abundances,
with mean values of [$m$/Fe] that have been described as the ``extreme"
Pop.~II metals mixture (see \tab{tab:abundances_pops}). The bMS stars
appear to have a higher iron content by about a factor of two than the
dominant more metal-poor stellar population, which has [Fe/H] $\approx
-1.7$.

The other, equally puzzling, CMD feature is the reddest MS that
is commonly referred to as MS-a.  We have shown that it is possible to
match the photometry of these stars remarkably well using isochrones for
the measured [Fe/H] value ($\approx -0.7$), provided that they also have
rather high helium and CNO abundances.  Indeed, it is possible that the
bMS and MS-a populations differ only in terms of their iron abundances;
i.e., that they are characterized by the same high values of $Y$ and the
``extreme" [$m$/Fe] ratios.  If this suggestion is correct, then all of
the stellar populations in \ocen\ would appear to have similar ages
(within $\approx$ 1 Gyr given the chemical abundance uncertainties).
For the MS-a stars to be significantly younger than, in particular, the
most metal-deficient stars, the former would need to have much higher
[CNO/Fe] abundances than those implied by the ``extreme" mixture, since
the latter abundances coupled with an old age (13 Gyr) are already needed
to explain the very faint subgiant branch of the MS-a component.
%To significant reduction in this age estimate can be achieved only if the
%stars have much higher C$+$N$+$O abundances. 

We have further established, through simulations of AGB and super-AGB
stars, that both of these types of stars, when forming out of the
abundance mix which has been assumed for the first and dominant [Fe/H]
$= -1.7$ population, are capable of ejecting winds with the high
helium abundances that are needed to explain the CMD locations of the
bMS stars.  However, for the source of this material, we favour the
massive AGB stars with CO cores and the less massive super-AGB stars
with ONeMg cores. The wind ejecta of the most massive super-AGB stars
with core masses $\geq 1.245\msun$ are likely too enriched with N
relative to He. In any case, even though either of the AGB models can
eject matter with sufficiently high He abundances, it is not clear
that they can produce ejecta with much larger He abundances than the
upper limit that has been set from the CMD of the bMS ($Y\leq0.4$).
This imposes limits on the acceptable amount of dilution of the AGB
material with gas that may still be present in the cluster or accreted
from the intercluster medium in order to allow bMS stars to form with
the indicated high He abundance pattern. Similarly, stellar
populations with very low O abundances set tight constraints on the
amount of allowable dilution \citep{D'Ercole:11}.

The obvious question is then how to isolate the AGB ejecta from other
gas that could be present; i.e., how to form the bMS stars out of just the
ejecta from massive AGB stars instead of, for instance, the C-rich ejecta
of somewhat lower-mass AGB stars (or possibly the Fe-rich ejecta of the
more massive stars that explode as supernovae). The observationally
established central concentration of stars with the ``extreme" abundance
mix \citep{johnson:10} may very well be an important clue.

\subsection{A scenario for the formation of multiple populations in
  globular clusters}
We need to combine the stellar evolution results described thusfar with
additional information in order to propose a general scenario for the
formation of multiple populations, such as the bMS and MS-a/RG-a 
components of \ocen.  In particular, it is necessary to consider the
early evolution of globular cluster progenitors in the gravitational 
potential of the host galaxy in combination with the cooling properties
of AGB ejecta.  To be sure, several aspects of this scenario have already
been suggested and elaborated upon elsewhere, as pointed out below.

\subsubsection{The progenitor of \ocen\ and the orbit in the Galactic potential}
\label{subsec:orbit}

The present-day orbit of \ocen\ consists of a super-position of several
oscillation components (\abb{fig:ocen-orbit}), which suggests a complex
capture mechanism that may very well have involved additional merging
components. According to the backwards in time integrated orbit 
\ocen\ has passed through the Galactic disk 13 times over the
past \natlog{5}{8}\jahre\ (\abb{fig:ocen-orbit}), implying an average time
interval between Galactic plane passage events of $\sim40$ million years, with some
variance. This estimate is approximate because of the uncertainty
associated with the present-day velocity of \ocen, and furthermore, the time
interval may have shortened significantly since the capture of \ocen's
progenitor, which may have been a dwarf galaxy
\citep{bekki:03,boeker:08,marcolini:08}.
\begin{figure}
%\plotone{Figures/OmegaCenOrbit.pdf}
\plotone{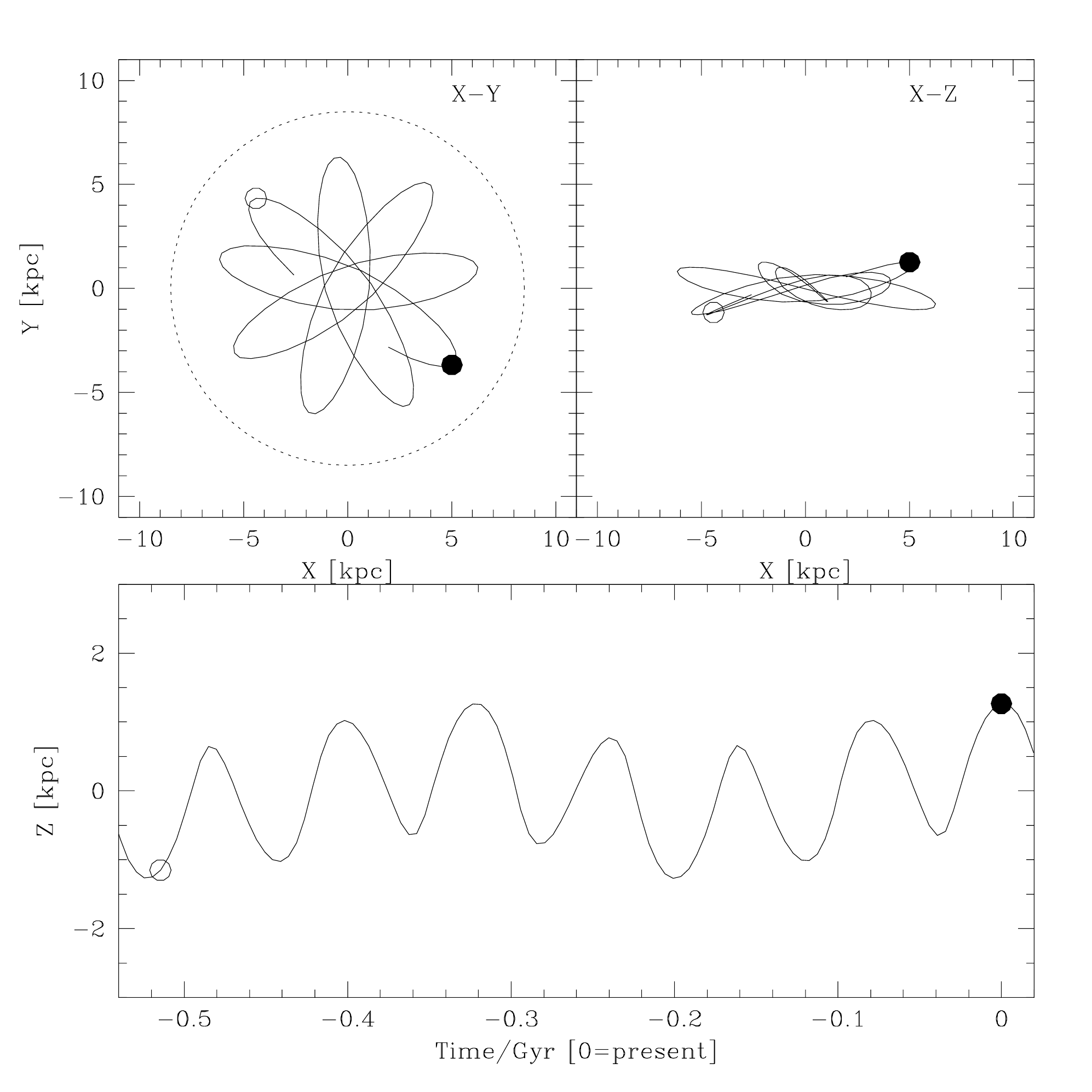}
\caption{
The orbit of $\omega$Cen integrated assuming the present-day
coordinates and positions given by \citet{Dinescu1999} and the Galaxy
model described in \citet{Michel-Dansac2011}. The orbit is computed by
integrating backwards in time (i.e., after changing the sign of the
present-day velocity) and shown for the past $500$ Myrs. The solar
circle is indicated, at $R_\odot=8.5$ kpc, by a dotted line. Note that
$\omega$Cen crosses the disk of the Galaxy roughly every $30$-$50$
Myrs.} 
\label{fig:ocen-orbit}
\end{figure}

In this primordial scenario of globular cluster formation \citep{padoan:97},
\ocen\ would have been much more massive \citep[e.g.,\ by a factor of
$\sim 25$ according to][]{bekki:03} when it was accreted, and it
would have lost substantial fractions of its stellar and halo mass
through tidal shocks at each pericentric passage.  This process has
been demonstrated through, for example, the $n$-body simulations by 
\citet{penarrubia:08} of dSph galaxies orbiting in the potential of
galaxies like the Milky Way.  These computations show that, depending
on the orbital parameters (specifically the apocentric/pericentric
ratio), within the first couple of pericentric passages (i)
$\sim\,90\%$ of the dark matter halo is lost before any significant
loss of stars occurs and (ii) mass is lost through outside-in
``onion-peel" stripping.  Through such tidal mass loss and evaporation
episodes, it is quite conceivable that a much more massive progenitor
would have ended up in \ocen's present state.

For globular clusters, Galactic plane passages have long been associated
with the ram pressure stripping of the gas content of globular clusters
\citep{Tayler1975}. We therefore assume that dark matter, stellar
and gas components of the merging progenitor of \ocen\ will be lost in
varying proportions \citep[as have, e.g.][]{valcarce:11,DAntona:2011dm}, that will depend on the exact mass and orbit as
well as the state of the Galaxy at the time of each interaction. The
outcome would in any case be a complete purging of the gas in Galactic
plane passages.

\subsubsection{Supernova feedback and purging}
\label{sec:SNfeedback}

Studies of supernova feedback in the context of the formation and
evolution of galaxies show that star formation correlates with the
mass of the galaxy and anti-correlates with the energy of the
supernova \citep{scannapieco:08}. Supernova feedback in low-mass
systems leads to more severely limited and bursty star formation. If
the progenitor of \ocen\ had low enough mass in its early history,
supernova feedback could have caused the purging of all of its gas.
Indeed, an alternative to the Galactic plane passage purging is gas
purging through supernova from stars with initial masses that bracket
the AGB mass range \citep{d'ercole:08,Charlie2011a}: stars more
massive than super-AGB stars explode as type II supernovae, while
those near the lower mass limit for massive AGB stars may lead to
prompt (single-degenerate) supernovae of type Ia.  In each of these
cases, depending on the mass of the stellar system, SN explosions
could purge the gas from it.

The cluster would, however, somehow need to be able to retain some of the iron
from SN (Ia ?) explosions at later times in its evolution in order to 
enrich the gas out of which the relatively metal-rich MS-a population
formed \citep{pancino:02,origlia:03,pancino:11}.  Similarly, at the
epoch when stars were evolving into massive AGB stars, the potential
well of \ocen\ likely has to be deep enough for this system to retain
the Fe ejecta from the small range of initial masses corresponding to
the lowest mass type II SN, which have the lowest explosion energies.
Most of the Fe that was produced by SN from more massive progenitors
would have been blown out of the GC along with any intra-cluster
gas.\footnote{In this picture, the cluster would have also expelled
  most of the wind material from rapidly rotating massive stars, which
  was proposed by \citet{decressin:07} to be the production site of
  the He-enriched material that went into the formation of such
  extreme populations as the bMS in \ocen. However, it has been
  shown by \citet{romano:07} that these special massive stars cannot,
  by themselves, account for the required abundance patterns of the bMS.}  

\citet{d'ercole:08} have pointed out that, if supernova from the
electron-capture core collapses of the ONeMg cores of super-AGB stars
have significantly lower explosion energies \citep{dessart:06}, the
(possibly) Fe-rich ejecta of those SN may be more easily retained by
the cluster.  This may be a viable alternative explanation of the small
increase of the iron abundance in the bMS stars compared to the measured
[Fe/H] of $\approx -1.7$ in the dominant initial stellar population.
\citet{marcolini:07} have investigated such aspects of the
evolution of \ocen\ through 3D hydrodynamic simulations of an isolated
progenitor system that was assumed to be a dSph galaxy. They emphasized
the differences between the SN II and Ia effects on the gas polution
and the possibility of inhomogeneous enrichment of gas with SN Ia
ejecta, but they admitted that tidal interactions with a host galaxy
may have additional important consequences. 

In any case, there are two, possibly complementary, processes available 
for the purging of all gas from the cluster --- SN purging or periodic
ram pressure/tidal shock purging through Galactic plane or pericentric
passages. Which of these mechanisms dominates will depend on the detailed
dynamical evolution of the system. In \kap{sec:scenario} we will explain
why we favor the purging of gas via Galactic plane passages in the
case of \ocen\ because of the unique effects that this process can
potentially have on the star formation and chemical evolution histories
of such a system.

\subsubsection{The Cooling of the Gas Ejected by AGB Stars}
\label{subsec:cool}

Cooling flows from AGB ejecta have been investigated by \citet{d'ercole:08},
on the assumption of a cooling function for the solar metallicity
by \citet{rosen:95}. However, the ejecta of the first generation AGB stars
have a much lower metallicity, and therefore one would expect that the
cooling of this material would be much less efficient.  Since the 
cooling flow of AGB ejecta to the center of a system like \ocen\ is
an important ingredient in most scenarios \citep[ours, and those 
by, e.g.,][]{d'ercole:08,D'Antona2011,bekki:11} for the origin of the
He-rich second generation (see \kap{sec:scenario}), it is worthwhile to
examine the cooling properties of a gas having the ``extreme" metals
mixture, and how it differs in comparison with that predicted for
for ``standard'' (i.e., typical) Pop.~II abundances. 

While the actual abundances in the \ocen\ bMS stars may be somewhat
milder than in our ``extreme'' abundance mix, those abundances out of
which the stars formed may have been subject to some limited amount of
dilution (cf.\ \kap{sec:disc-main-results}). Any assumption of
dilution requires the AGB ejecta to be more extreme than the bMS stars
in \ocen. What exactly the abundance of the cooling flow is will
depend on when the dilution takes place, something that is difficult
to know. In any case, it is appropriate to consider the cooling
properties of material with``extreme'' composition as representative
of what the AGB ejecta may look like.
\begin{figure}
%\plotone{Figures/ocenf4.pdf}
   \includegraphics[width=0.48\textwidth]{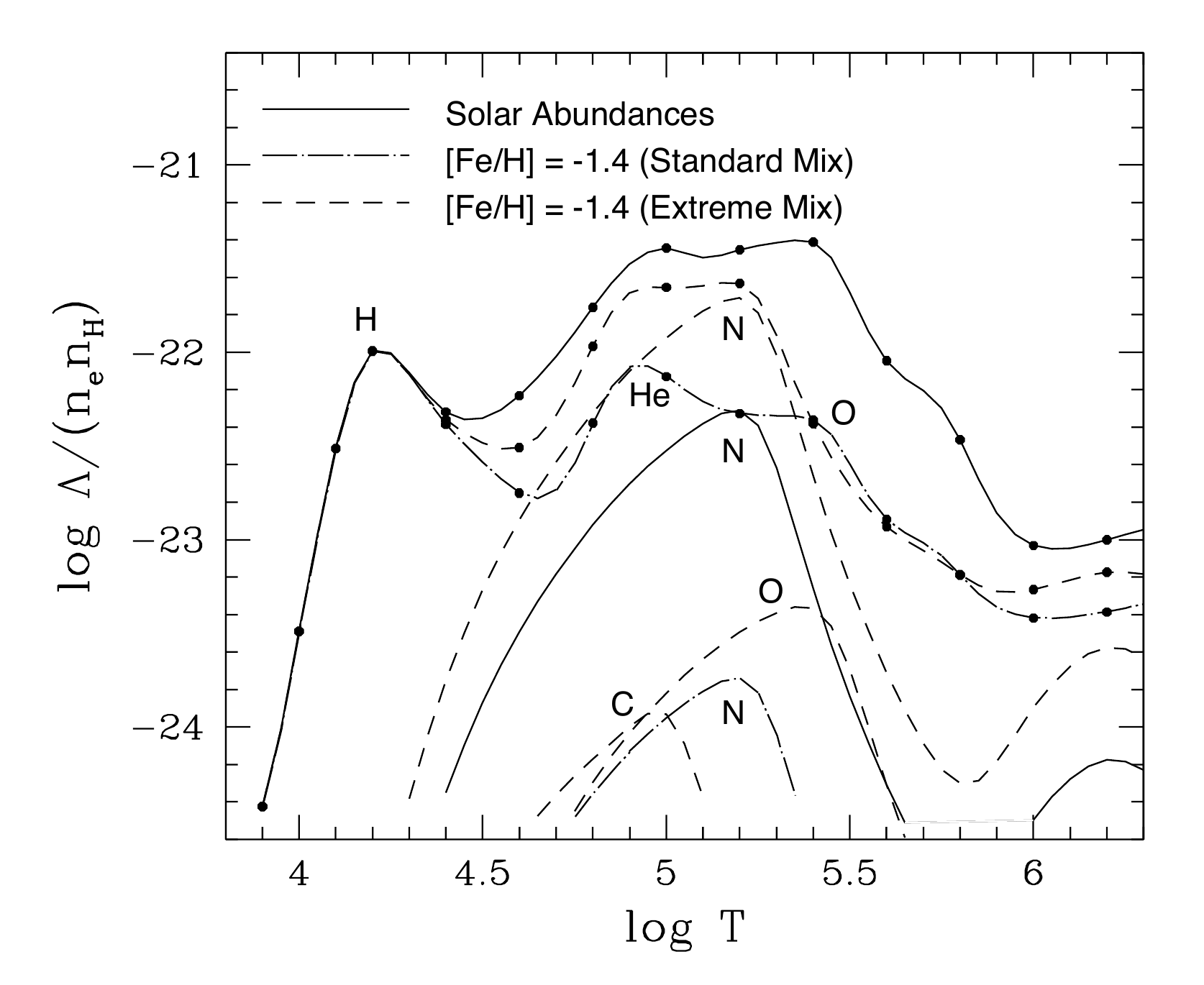}
\caption{The radiative cooling coefficient for a gas having the
  indicated chemical abundances (also see the text).  The loci with
  filled circles superimposed represent the sum of the contributions
  to $\Lambda/(n_en{\rm H})$ from H, He, C, N, O, and Ne, as well as
  free-free emission (which dominates at $\log T \gta 5.8$).  For a
  sake of clarity, only a few of the contributions to the total
  cooling curves from individual elements have been
  plotted. }
\label{fig:fig4}
\end{figure}

In the limit where the photoionization caused by the stellar radiation
field can be neglected; i.e., when collisional ionization conditions
apply, the radiative cooling coefficient (in
erg$\,$cm$^{-3}\,$s$^{-1}$) varies as a function of $\log\,T$
approximately as shown in Fig~\ref{fig:fig4} for three different
heavy-element mixtures.  This plot was generated using the computer code
developed by VandenBerg (1978) to model the outflows of gas from
present-day globular clusters to try to explain why the gas that is
ejected into the interstellar medium (ISM) through normal
(low-velocity) mass-loss processes is not observed.  (When these
systems pass through the Galactic disk, any gas that might have
collected is expected to be swept out by the ram pressure exerted
by the local ISM, but gas should accumulate to detectable levels over
the rest of their orbits in at least the most massive systems.)  If
photoionization is unimportant, then the ionization state of the gas
can be determined simply from the condition that the number of
collisional ionizations per unit time for any ion is exactly equal to
rate of electron recombinations, which is independent of the density.
Once the ionization state has been calculated, the radiative energy
losses due to free-free and free-bound transitions, and from the
transitions to lower energy levels from high-excitation states that
were populated by recombination or inelastic electron collisions, can
be determined (see the VandenBerg paper for a detailed description of
the relevant physics).

Since the energy loss rates due to recombination radiation and
collision-induced line emission are proportional to $n_en_{Z,z}$,
where $n_e$ is the electron density and $n_{Z,z}$ is the number
density of atoms having an atomic number $Z$ and an ionic charge $z$,
the radiative cooling coefficient $\Lambda/(n_en_{\rm H})$ can be
derived for any mixture once the fractional ionization, $n_{Z,z}/n_Z$,
has been calculated and the abundances of each element are specified
relative to hydrogen (i.e., $n_Z/n_{\rm H}$).  The solid curves in
Fig.~\ref{fig:fig4} assume the solar metal abundances given by
\citet{asplund:09}, while the dot-dashed and dashed curves give, in
turn, the temperature dependence of $\Lambda/(n_en_{\rm H})$ for an
\citet{asplund:09} mixture having [$m$/H] $= 0.4$ for all of the
$\alpha$-elements (``standard'' composition in \tab{tab:abundances_pops})
and then scaled to [Fe/H] $= -1.4$ and for the ``extreme'' mix from
\tab{tab:abundances_pops} scaled to [Fe/H] $=-1.4$. In these latter
cases, the adopted mass fractions of helium correspond to
$Y = 0.25$ and $Y = 0.38$, respectively.  (Since these results may be
of some use to future investigations that take into account the cooling
of the gas, the total cooling coefficients, as represented by the loci
with filled circles attached to them, have been listed as a function of
$\log\,T$ in Table~\ref{tab:tab4}.)
\begin{deluxetable}{cccc}
\tablecolumns{4}
\tablecaption{Radiative Cooling Coefficient: $\log\,(\Lambda/n_en_{\rm H}$)
 \label{tab:tab4}}
\tablewidth{0pc}
\tablehead{ 
\colhead{} &\colhead{[Fe/H] $= 0.0$}  & \multicolumn{2}{c}{[Fe/H] $= -1.4$} \\ 
\colhead{$\log T$} & \colhead{}&
\colhead{ Standard Mix} & \colhead{Extreme Mix }}
\startdata  
 3.80 & -24.694 & -24.694 & -24.694 \\
 3.85 & -24.626 & -24.626 & -24.626 \\
 3.90 & -24.425 & -24.425 & -24.425 \\
 3.95 & -24.012 & -24.013 & -24.013 \\
 4.00 & -23.489 & -23.490 & -23.490 \\
 4.05 & -22.972 & -22.973 & -22.973 \\
 4.10 & -22.513 & -22.514 & -22.514 \\
 4.15 & -22.162 & -22.163 & -22.163 \\
 4.20 & -21.992 & -21.994 & -21.993 \\
 4.25 & -22.005 & -22.009 & -22.009 \\
 4.30 & -22.110 & -22.121 & -22.118 \\
 4.35 & -22.228 & -22.257 & -22.248 \\
 4.40 & -22.319 & -22.386 & -22.360 \\
 4.45 & -22.359 & -22.493 & -22.433 \\
 4.50 & -22.353 & -22.589 & -22.482 \\
 4.55 & -22.308 & -22.679 & -22.514 \\
 4.60 & -22.232 & -22.752 & -22.509 \\
 4.65 & -22.134 & -22.783 & -22.453 \\
 4.70 & -22.022 & -22.736 & -22.337 \\
 4.75 & -21.896 & -22.589 & -22.168 \\
 4.80 & -21.761 & -22.377 & -21.968 \\
 4.85 & -21.631 & -22.183 & -21.788 \\
 4.90 & -21.530 & -22.079 & -21.682 \\
 4.95 & -21.466 & -22.076 & -21.651 \\
 5.00 & -21.443 & -22.130 & -21.653 \\
 5.05 & -21.469 & -22.202 & -21.654 \\
 5.10 & -21.493 & -22.266 & -21.644 \\
 5.15 & -21.482 & -22.308 & -21.629 \\
 5.20 & -21.454 & -22.328 & -21.632 \\
 5.25 & -21.429 & -22.338 & -21.713 \\
 5.30 & -21.415 & -22.342 & -21.912 \\
 5.35 & -21.402 & -22.343 & -22.161 \\
 5.40 & -21.412 & -22.362 & -22.381 \\
 5.45 & -21.496 & -22.443 & -22.559 \\
 5.50 & -21.680 & -22.602 & -22.712 \\
 5.55 & -21.889 & -22.771 & -22.836 \\
 5.60 & -22.046 & -22.892 & -22.931 \\
 5.65 & -22.140 & -22.968 & -23.002 \\
 5.70 & -22.205 & -23.022 & -23.060 \\
 5.75 & -22.298 & -23.089 & -23.120 \\
 5.80 & -22.467 & -23.190 & -23.187 \\
 5.85 & -22.677 & -23.293 & -23.244 \\
 5.90 & -22.857 & -23.363 & -23.275 \\
 5.95 & -22.974 & -23.402 & -23.280 \\
 6.00 & -23.031 & -23.419 & -23.266 \\
 6.05 & -23.050 & -23.423 & -23.242 \\
 6.10 & -23.046 & -23.417 & -23.215 \\
 6.15 & -23.028 & -23.405 & -23.190 \\
 6.20 & -23.001 & -23.388 & -23.175 \\
 6.25 & -22.972 & -23.368 & -23.173 \\
 6.30 & -22.947 & -23.348 & -23.183 \\
\enddata
\end{deluxetable}

For the sake of clarity, just a few of the contributions to the total
cooling rates, due to the individual metals, are shown.  Nitrogen is of
particular interest in view of the wide range in its abundance in the
three mixtures, and it is quite evident that the cooling due to this
element in the ``extreme" [Fe/H] $= -1.4$ mixture is much greater than
that predicted for the other two cases (including, in particular,
solar abundances).  It is also much greater, by factors of $\gta 60$,
than the contributions due to C and O because the latter are much less
abundant.  As helium is mainly responsible for the bump in the
radiative cooling coefficient at $\log\,T\approx 4.9$, the effects of
increasing $Y$ from 0.25, in the ``normal" mix, to 0.38, in the
``extreme" mix, are clearly significant.  The main point of
Fig.~\ref{fig:fig4} is that a gas with high N and He
abundances will cool much more efficiently than a gas having a more
normal mix of metals and a helium abundance that is closer to the
primordial value.  Indeed, the total radiative cooling coefficient for
the former approaches that for a gas having solar abundances (at least
at $\log T \lta 5.3$).

The presence of an intense stellar radiation field, such as that
produced by extremely hot ($\sim 50,000$ K) $uv$-bright stars,
would heat the gas and ionize many of the atoms and ions that are
responsible for radiative cooling, and thereby hinder or prevent
the infall of gas towards the cluster center. However, such stars,
which represent a much more significant source of heating than large
populations of blue HB stars (see VandenBerg 1978), are quite rare
($\lta 1$ star is expected in present-day GCs at any given time
although this number is expected to have been larger in the past).
Consequently, it is difficult to describe the interplay between the
radiation field and the material shed by AGB stars, without
performing detailed simulations of globular clusters when they were
younger and much more massive than they are now. 

It will be interesting to see what impact these more realistic cooling
properties will have on simulations that include the cooling flow of
AGB stars \citep[as well as that of primordial matter,
e.g.][]{d'ercole:08}. Considering the ``extreme'' abundance mix as an
upper limit of the boost in cooling efficiency over the ``standard''
mix, the cooling efficiency of realistic AGB ejecta are smaller by a
factor $\apgeq 4$ compared to solar cooling curves employd be those
simulations.  In any case, we do assume in the following that AGB
ejecta will cool and accumulate in the cluster center \citep[see
also][who comes to the same conclusion]{bekki:11}.
 
\subsubsection{A possible scenario}
\label{sec:scenario}
The lifetime of the $7$ and $5\msun$ stellar evolutionary models is
$4.6$ and $\natlog{9.1}{7}\jahre$, respectively
(\abb{fig:mini-starage}).  The similarity of the time interval between
successive Galactic plane passages of \ocen\ and the time it would
take super-AGB and massive AGB stars to expel their He- and N-enriched
and O-depleted wind ejecta could very well conspire in such a way
that, by the time that the massive- and super-AGB stars are about to
eject their envelopes, the wind material from more massive stars has
been entirely, or mostly, cleared out of the system and the AGB winds
are released into an empty, or nearly empty, cluster. Stars with an
extreme abundance mix would subsequently form, but the star formation
process would be terminated when the remaining gas is cleared out
during the next passage through the plane.  This would isolate the AGB
yields and enable formation out of pure AGB ejecta.

The scenario for the origin of multiple populations starts with the
progenitor evolution of \ocen.  The proto-cluster, possibly a dwarf
galaxy (\kap{subsec:orbit}), would have a fully populated initial-mass
function (IMF) when it merges with the Galaxy.  Star formation may
have going on at a low level since its formation, corresponding to the
low-metallicity tail below the dominant (first) generation at
[Fe/H]$\sim-1.7$ in \ocen. But somehow, we assume that the dominant
stellar content of the \ocen\ progenitor would be the first
generation, and it would have been produced by a star formation burst,
maybe triggered by the tidal interactions associated with the complex
capture processes indicated by the present-day orbit
\citep[\abb{fig:ocen-orbit}, \kap{subsec:orbit}; see
also][]{bekki:03}. Assuming that \ocen\ is then in an orbit fairly
similar to the one shown in \abb{fig:ocen-orbit}, the \emph{first}
Galactic plane passage gas purging could occur $\lta 40$ million years later.
This would remove all of the remnant gas ejecta from stars more
massive than super-AGB stars.

Over the next $50$--$90$ million years, until the next Galactic plane
passage, AGB stars and possibly super-AGB stars would eject winds into
the empty cluster, with
abundance patterns that have helium and oxygen abundances that closely
resemble those attributed to the bMS stars in \ocen\
(\kap{sec:agb}). These ejecta would initially have the same spatial
distribution througout the cluster as the AGB donor stars, but
radiative cooling processes would cause the gas to sink preferentially
towards the center of the cluster, as disccussed in \kap{subsec:cool}.

According to \citet{bekki:11}, the accumulation of gas and the
subsequent formation of the second generation of stars will be fully
under way after $\sim 3$ million years, and be largely completed in
the central core within $\lta 13$ million years. This leaves enough
time until the next Galactic plane passage to fully populate a
second-generation IMF.  

Bekki suggests that the second-generation IMF may be bottom heavy in
order to explain the high ratio of second- to first-generation stars.
However, as mentioned in \kap{subsec:orbit} and noted by Bekki, among
others, further mass is lost from the system through subsequent
pericentric passages from the outside-in (see the discussion in
\kap{subsec:orbit}). In fact, the fraction of first generation stars
that has to be lost (from the outside-in) in order to match the
observed bMS star fraction can be easily estimated for a given mass
range of intermediate mass stars that release their winds between
Galactic plane passages. This mass range will of course depend on the
(not accurately known) orbital parameters of \ocen\ at that
time. Based on the calculated model grid described in
\kap{sec:inimass} we adopt a lower limit for this mass range as
$M_\mem{min} = 4$ -- $5\msun$ by the requirement to produce large He
and N enhancements, and $M_\mem{max} = 6.8$ -- $7.7\msun$ taking into
account possibly prohibitive dredge-out (\kap{sec:deep2dup}) and
excluding the high-mass regime of core-collapse supernova. With this
mass range assumptions the IMF of \citet{kroupa:01} implies that
intermediate mass stars will eject $5.2\pm2.0\%$ of the mass of the
first generation stars (assuming $ 20\%$ of the initial stellar mass
will end up in the white dwarf). The number of bMS stars has been
projected to be $24$ -- $35\%$ of the first generation stars
\citep{bedin:04,sollima:07b}. These numbers imply that in order to account for
the observed fraction of bMS stars $78.3\pm8.3$ -- $85.1\pm5.7\%$ of first generation
stars must have been lost from the outside (assuming a normal IMF).
This seems to be reasonable in our scenario (cf.\ \kap{subsec:orbit}),
considering also that the central location of the bMS population
implies that it contributes only marginally to the mass that is
lost from the cluster through dynamical interactions. We therefore
assume that the second generation IMF is normal.

Then, the second generation will also include massive stars.  The
associated supernova will explode in the very center of the still
relatively massive cluster, which is the most favorable location for
the retention of SN ejecta \citep{marcolini:07}.  Depending on when
these SN occur with respect to the timing of the \emph{second}
Galactic plane passage, the ejecta from the most massive
second-generation supernova may be lost during the passage through the
disk, but afterwards, it seems plausible that Fe-rich ejecta from the
lowest-mass SNe could be retained in the core region. Further down the
IMF, super-AGB stars and possibly the most massive AGB stars with CO
core could again contribute to what would be the third generation.
This would again form in the cluster core before the \emph{third}
Galactic plane passage would terminate this last major star formation
episode. This third generation would have a substantially higher iron
abundance from second-generation SN, compared to that of the bMS, and
it would have enhanced He and CNO.  Furthermore, this third
generation, which may be the RGB-a/MS-a population discussed in
\kap{sec:cmd}, would consist of significantly fewer stars than the
second generation. As before, this star formation event would end when
the cluster makes its \emph{fourth} passage through the Galactic plane
and the remaining gas is purged. After that diminishing numbers would
make it impossible to discern further populations in the cluster core.

Star formation would not be impossible in other parts of the cluster
after this \emph{fourth} Galactic plane passage, but it is possible
that, by this time, the mass of the cluster has decreased to the
extent that it is no longer possible to retain the ejecta from
supernovae.  In this case, the total cluster wind ejecta may now,
$\geq \natlog{2}{8}\jahre$ after the initial star formation burst, be
dominated by the increasing number of intermediate-mass ($\sim 3
-4\msun$) first-generation stars that are ending their lives.  The
ejecta would again be $s$-process rich, but also C-rich.  Stars
forming out of these ``late" wind ejecta would have rather low [Fe/H]
since they are formed out of first (and possibly second) generation
star ejecta. This may explain why the \spr\ enhanced stars do not show
the same central concentration compared with the He-enriched stars,
and why the \spr\ enhancements seem to be increasing already at
relatively low [Fe/H], not much higher than that of the first
generation. As a matter of fact, the data of \citet[][Fig.\,6 \&
7]{marino:12} indeed shows two La-enrichment sequences, one
belonging to what is identified as the 'first generation' that
correlates with C, and one that correlates rather with N, and
associated with the second generation. In our scenario, the C-rich,
low [Fe/H] and high La-abundance stars would in fact be the fourth
generation (see \kap{sec:final-ONa} for more discussion on this point).

As discussed in \kap{sec:SNfeedback}, some workers
\citep[e.g.][]{d'ercole:08,Charlie2011a} favour the idea that the
purging of gas from clusters is mostly due to supernova.  Massive-star
SN II explosions would clear out the gas until a lower mass limit is
reached where either SN IIs no longer occur, or the explosions are so
weak that they are incapable of removing all of the gas. Whether or not
SN are effective in purging the gas sufficiently to enable the formation of
a second generation out of the isolated ejecta of stars from a particular
mass range will not only depend on the total mass of the system and the
energy of the supernova, but also on their location within the cluster
(progenitor). While SN purging clearly seems to be a viable mechanism
in some (possibly less massive) clusters, our proposed scenario appears
to be capable of explaining the detailed properties of the very unusual
case of \ocen, such as the small difference in [Fe/H] between the first
and second stellar generations, and the larger difference in metallicity
between the second and third generations.  Moreover, it should also
work for other clusters that are massive enough to retain SN ejecta.

\begin{table*}
\caption{ Summary of the four generations identified in the Galactic
  plane passage gas purging model for \ocen, and how they match with
  the cluster analysis of \citet{Gratton:2011kr} and with the
  observational properties reported in DOrazi:2011jf and \citet{marino:12} (see text for
  details, in particular \kap{sec:final_sum}).
  \label{tab:gppgpm}}
\begin{tabular}{p{0.08\textwidth}p{0.07\textwidth}p{0.28\textwidth}p{0.06\textwidth}p{0.24\textwidth}p{0.16\textwidth}}
\hline\hline
generation&parent generation&expected abundance markers&
\citet{Gratton:2011kr} group& corresponds to
features reported in \citet{marino:12}& Leiden star numbers \citet{DOrazi:2011jf}\\
\hline
first       & proto-cluster& low Fe; normal He, La; high $\alpha$,
high (from r-process) Eu & \#4, \#6 &  & 16015, 37247, 33011,  \\
second, bMS & first        & low O, intermediate Fe; high He, N, Na,
n-capture  & \#3, \#2a &    low O, high Na (most extreme?) for medium
[Fe/H] & 41039, 46092, 34029 \\
third, MS-a/RGB-a & second       & high Fe, C+N+O; higher N, He, n-capture
&\#2b  &  highest Na \& La for highest [Fe/H]&  60066, 60073, 34180,
48323, 54022\\
fourth & first (second)   & low (medium) Fe; normal He, medium O, Na, high C,
n-capture  & \#5, \#1 & O, Na, N, C - La relation of O-poor/Na-rich
group & 44462 \\
\hline\hline
\\
\end{tabular}
\end{table*}

\section{Conclusions}
\label{sec:conclusions}

\subsection{Implications of the Galactic plane passage gas purging model}
As \citet{d'ercole:08} put it, He-rich populations (like the bMS in
\ocen) are only the tip of the iceberg of the phenomenon of second (or
third, etc.) stellar populations. While our scenario of periodic gas purging
events caused by the passages of clusters through the Galactic plane
appears to be able to explain quite naturally the origin of the
helium-rich bMS in \ocen\ (and possibly the more Fe-rich MS-a/RGB-a
population, as well), one can readily imagine that this simple picture,
if applied to clusters having a wide range of progenitor histories
(e.g., masses and orbits),
%and possibly in repeated superposition (as eluded to in the
%previous paragraph),
may, in fact, generate quite a variety of realizations in the globular
clusters that we observe today.  Since each population would be
associated with the ejecta arising from stars within one, or very few,
relatively narrow initial mass ranges, isolated by successive Galactic
plane passages that clear out any gas that had accumulated since the
last passage, any two (or more) populations that are present in
different clusters should not be identical, though possibly similar.
The contributing segments of the IMF are statistically selected
according to the progenitor orbit and the merger history.

However, in our scenario, Galactic plane passage intervals of $40$
to $100$ million years do favor the formation of a second, helium-rich
population. If the delay is $40$ million years \citep[as also found
by][]{marino:12}, upper and lower
mass limits can both be imposed by the Galactic plane passages.
For time intervals as large as $100$ million years, only the lower
mass cut-off for the progenitor stars of the second generation is set
by the Galactic plane passage, while the higher mass cut-off would be
due to SN purging. In fact, this case is similar to the model suggested
by \citet[][Sect.\,3.1]{d'ercole:08}, who adopt $100$ Myr for their
parameter $\Delta t_\mem{f}$. They do not mention what sets this time
interval, aside from noting that, if $\Delta t_\mem{f}$ was longer,
SN Ia and C-producing AGB stars would contribute to the second-generation
star formation (see the above discussion). This is indeed why $\Delta
t_\mem{f}$ should not be longer if the desired outcome is a helium-rich
second generation, but we suggest that the main reason is that the Galactic
plane passages which clear out the gas and terminate star formation occur
every $\lta 100$ Myr.

In summary, the simple principle of periodic Galactic plane passage
purging in combination with low-velocity winds from massive AGB stars
AGB wind predictions and their preferential cooling properties, may be
able to account for the bMS abundances including its central
concentration, the MS-a/RGB-a abundance patterns as well as the
homogeneous distribution of and wide range of [Fe/H] observed in
s-enhanced stars.

\subsection{O-Na anti-correlation, n-capture element abundances, and other
  observed properties of \ocen}
\label{sec:final-ONa}
One of the \citep[possibly defining,][]{Carretta:2010il} features of
globular clusters is the more or less complete presence of the O-Na
anticorrelation. How does it fit in with the Galactic plane passage
gas purging model? The key question is whether (a) the
anti-correlation occurs within a sub-population, or (b) it is the
superposition of the rather distinct O-Na abundances of the present
sub-populations. The answer will depend on how the latter are
identified. If one uses only the [Fe/H] abundance
\citep[e.g.][]{DAntona:2011dm} one may indeed combine N-rich/C-poor
with C-rich/N-poor stars into one group \citep[][Fig.\,2]{marino:12}
and conclude that abundance anti-correlations are present even within a
sub-population.

This view point may motivate, or even require as an explanationof the
Na-O anti-correlation, a dilution scenario \citep{DErcole:2011jca}
which assumes that, while second-generation stars form in the cluster,
unprocessed pristine gas is accreted from outside of the
cluster. However, if option (b) is in fact the case, for example, 
if the sub-population identification purely by [Fe/H] is not
entirely accurate, then an alternative interpretation of the
anti-correlation is possible in which dilution may not play an
important role.

Option (b) may also be favoured by observations of rather uniform Al
abundances in the SGB-a population (in our scenario the third
generation, see \tab{tab:gppgpm}) by \citet{pancino:11}, suggesting
that no anti-correlation is present in this sub-population. Although
we have tried to resist the temptation in this paper to apply the
Galactic plane passage gas purging model to other clusters we note
that \citet{Carretta:2012jp} found that the anti-correlations
involving Al, Mg, Na and O in NGC\,6752 -- another GC for
which multiple populations possibly including large He-enrichements have
been found -- manifest themselves in a rather discrete fashion, in
which different levels of enhancement and depletion cluster around
discrete values that can be associated with individual sub-populations,
rather than a continuous distribution that would be sugggested by a
pure dilution mechanism.

The Galactic plane passage gas purging scenario implies that two
distinct populations may have the same [Fe/H] abundance, and therefore
that a collection of stars with the same [Fe/H] may not necessarily
belong to the same population. As a matter of fact, many possible
elemental markers if used by themselves may lead to degenerate
grouping of stars. In order to identify those stars that most likely
represent a coevally formed sub-population, several elemental markers
should be combined. Such a ``group'' analysis, taking into account
four abundance features simultaneously, has been performed by
\citet{Gratton:2011kr}. The resultant sub-population identification is
indeed rather suggestive of option (b) mentioned in the previous
paragraph (superposition of the distinct O-Na abundance markers from
sub-population). Of course, there is always a concern that the
particular choice of group-finding criteria biases the process. For
that reason the sub-population identification by
\citet{Gratton:2011kr} may be evolve in the future when more
observational data are added to this kind of analysis. In any case,
this approach seems to be an improvement over just using [Fe/H] to
identify sub-populations. As a consequence, we have no compelling need
for the dilution mechanism in our model for \ocen.

Our scenario has a few more implications -- some of which are
summarized in \tab{tab:gppgpm} -- that we would like to briefly
discuss. It offers an alternative interpretation of the star
formation time-scale determination of \citet{DAntona:2011dm} that was
based on the assumption that Fe in the MS-a/RGB-a populations
originates in SN Ia. In our scenario, Fe for the \emph{third} generation
(MS-a/RGB-a) comes from the lowest mass, \emph{second} generation SN II  after the second
Galactic plane passage (\kap{sec:scenario}), which would be in
agreement with the $\alpha$-element abundance patterns found by
\citet{Gratton:2011kr} for the most Fe-rich population labeled
\#2b. The time-scale limit from SN Ia would therefore not
apply. In addition, star formation can take place even after the
formation of this third generation from the late stellar winds of
low-mass first-generation stars. 

In fact, if we retain our chemical evolution time range
determination from CMD considerations ($\lta 1$ Gyr, \kap{sec:cmd}),
stars in the fourth generation can form from the slow wind ejecta of
first-generation stars down to $1.8\msun$. For stars with $\apleq
2.6\msun$ these ejecta would be C-rich (as well as La-rich from the main
\sprn). Because the first-generation donor stars are not centrally
concentrated, the fourth generation stars that form out of them would
also not be centrally located --- if the cooling efficiency
of lower-mass AGB ejecta are smaller, or because the mass of the
cluster is already smaller at this point. Indeed, \citet{johnson:10}
report no radial gradient for [La/Fe]. In any case, the
observations reported by \citet{marino:12} show that there are two
different La-enrichment sequences, one associated with increasing C
(O-rich/Na-poor group) and one associated with low O and high Na. In
our scenario the former represents the transition from first to
\emph{fourth} generation, while the
latter is associated with both the \emph{second} and, possibly for even more
extreme levels of enrichment, the \emph{third} generation.

If we interprete the correlation of C with La in the O-rich/Na-poor
group of \citet{marino:12} as the result of the third dredge-up in
$1.8$ -- $2.6\msun$ \emph{first} (and maybe \emph{second} slightly higher
[Fe/H]) generation stars, then their Fig.\,7 implies that O and
Na could also be the result of the third dredge-up. In fact, the slight
increase of O with [Fe/H] in the O-rich/Na-poor group has been
perceived to be quite a puzzle that implies an extra source of
O. However, the models of \citet[][Fig.\,6]{herwig:04a} for
$\mathrm{[Fe/H] } = -2.3$ do indeed predict $\mathrm{[C/Fe]} = 3.0$,
$\mathrm{[O/Fe]} = 1.8$ and $\mathrm{[Na/Fe]} = 1.3$ for the average abundance
in ejecta of a $2\msun$ model. The O comes from
the dredge-up of primary He-burning products that becomes appreciable at
these low metal abundances, and such O enhancements are indeed
observed in many CEMP stars that may carry the mass transfer signature
from genuine low-metallicity AGB stars \citep[for
example,][]{sivarani:06,Kennedy:2011db}. The Na in the $2\msun$ model
comes from \sprn\ in the He-shell flash convection zone, where the
neutron from the $\nezw(\alpha,\n)\mgfu$ reaction is captured again by
\nezw. The predicted enrichment levels suggest that a population which
forms out of such ejecta should have about twice the enrichment of C
compared to O and Na, which seems to be consistent with the
\citet{marino:12} data.

Star Leiden\, 44462 has been tentatively identified by
\citet{DOrazi:2011jf} as a mass transfer object in order to account
for the extremely high C abundance. Although radial velocity
measurements may support this possibility, an alternative
interpretation is that this star is part of the fourth generation
(\tab{tab:gppgpm}) which forms out of the slow winds of $1.8$ --
$2.6\msun$ first (or second for higher [Fe/H]) generation stars. In
fact, this star coincides very well with the C-La correlation sequence
shown in \citet[][Fig.\, 7, upper right panel]{marino:12}.

But the O-poor/Na-rich group of \citet{marino:12}, as well as all
but the lowest [Fe/H] stars in \citet{DOrazi:2011jf}, also show marked
n-capture enhancements. Since in these stars the C abundance is low,
the heavy elements cannot come from these lower-mass AGB
stars. Instead we have to consider higher-mass AGB stars as well
as super-AGB stars. The n-capture element predictions of
\citet{Karakas:2012kc}, e.g.\ their  $6\msun$, $Z=0.0001$ stellar
evolution model, are based mostly on the \nezw\ neutron source in the
He-shell flash convection zone (although some contribution from a
\cdr-pocket may be present as well). Clearly, the \spr\ models for
these low-metallicity intermediate-mass stars are quantiatively still
rather uncertain (cf.\ \kap{sec:increased_amlt}). However, the models
of \citet{Karakas:2012kc} do predict that higher mass AGB stars at
this metal content do produce \spr\ elements, possibly with
significant enrichment factors, and with a ratio of light (ls: Sr, Y,
Zr, Rb) to heavy (hs: Ba, La) \spr-elements that is higher than in the
ejecta of lower mass stars. Such a signature would qualitatively agree
with the n-capture abundances reported for the N-rich, intermediate-
(bMS, second generation) and high- (MS-a/RGB-a, third generation) Fe
group by \citep{DOrazi:2011jf}.

  In addition to the n-capture production by $\nezw(\alpha,\n)\mgfu$
  in the He-intershell in intermediate mass AGB stars, we mentioned in
  \kap{sec:inimass} the possible presence of \ipr-conditions in
    super-AGB models with CBM \citep{herwig:12b}, which would provide
    for another ``non-standard'' source of n-capture elements in stars
    that produce He-rich and O-poor populations in \ocen. Both of
    these sources would be responsible for the La-enhanced
    sequence of the O-poor/Na-rich group reported
    by \citet{marino:12}. 

    An important consequence of this discussion is the notion that the
    Na enhancements which we expect from AGB stars from $1.8\msun$ all
    the way up to the super-AGB stars at $\apleq 6.8$ -- $7.7\msun$ may
    always be expected to go along with n-capture element
    enhancements. This assessment is supported by the La-Na
    correlations of both the O-rich/Na-poor and the O-poor/Na-rich
    group of \citet[][lower-right panel of Fig.\,7]{marino:12}.

\subsection{Summary}
\label{sec:final_sum}
Obviously the observational identification of the four generations
that we have specified -- within the Galactic plane passage gas
purging model -- as the result of highly idealized processes, is
complicated by interference and superposition effects as well as
contributions from, e.g., supernova purging, turbulent and tidal
mixing, and cooling and mass loss flows, all of which are expected in
a real cluster. In addition, observational uncertainties may cause some
migration between observationally identified groups, which is in
addition to the principal difficulty of deciding on the criteria and
procedures that are used to group stars (as discussed at the
beginning of \kap{sec:final-ONa}). These complications impose a limit
to the accuracy that we can expect in how well observed properties can be
matched to the predictions of any scenario. The best that we can ask for at
this point is rather qualitative agreement, and with this goal in
mind we have summarized the alignment of the Galactic plane passage
gas purging model with some recently reported observational properties
of \ocen\ in \tab{tab:gppgpm}.

While most of the entries in \tab{tab:gppgpm} are based on our
discussion in the previous sections there is a noteworthy peculiarity
in the most metal-poor group (``first'' generation) which should
represent the genuine first generation stars. In \kap{sec:final-ONa}
we made the case that the observational data may support the
case of the superposition of separate sub-populations forming the
overall anti-correlation in \ocen. It seems that, in fact, in this
lowest [Fe/H] bin there are signs of an intrinisic anti-correlation
with at most an unclear signature of n-capture enhancement. Groups \#6
and \#4 of \citet{Gratton:2011kr} taken together display a significant
scatter in the [Na/O] ratio, while the lowest-metallicity bin of
\citet{DOrazi:2011jf} includes two very N-rich stars \citep[see
also][]{marino:12}. This possible substructure in the most metal-poor
population in \ocen\ cannot be explained by the Galcatic plane
passage gas purging model, and may be a relic of the proto-cluster
object.

\acknowledgements This paper was significantly improved and extended
as a result of the thorough and thoughtful report provided by the
referee, Santi Cassisi, and we are very grateful to him for the time
and effort that he put into his critique.  We thank Antonio Sollima
and Luigi Bedin for providing the photometry of $\omega$ Cen that has
been used in this investigation, as well as helpful comments.  We are
also grateful to Santi Cassisi for sending us his color transformation
tables applicable to ACS photometry, and to Leo Michel-Dansac, who
kindly assisted with the orbital integration of \ocen.  Thanks go, as
well, to Kim Venn, David Hartwick, Pavel Denisenkov, and the entire
``stars group" at the University of Victoria for many stimulating
discussions concerning various topics of stellar evolution, globular
clusters, and stellar populations in general. This work has been
supported by the Natural Sciences and Engineering Research Council of
Canada through Discovery Grants to FH, DAV and JFN.  This research has
also been supported by the National Science Foundation under grants
PHY 11-25915 and AST 11-09174.

%\bibliography{astro}
%\end{document}

\end{document}